\RequirePackage[final]{graphicx}
\documentclass[notocbib, aps, superscriptaddress, prl, twocolumn, footinbib]{revtex4-1}

\usepackage[utf8]{inputenc}
\usepackage{amsmath}
\usepackage{amsfonts}
\usepackage{mathtools}
\usepackage{graphicx}
\usepackage{hyperref}
\usepackage{fixme}
\usepackage{color}
\usepackage{braket}
\usepackage[compat=1.1.0]{tikz-feynman}
\usepackage[normalem]{ulem}

\newcommand{\nocontentsline}[3]{}
\newcommand{\tocless}[2]{\bgroup\let\addcontentsline=\nocontentsline#1{#2}\egroup}

\newcommand{\beginsupplement}{%
        \setcounter{section}{0}
        \renewcommand{\thesection}{S\Roman{section}}%
        \setcounter{equation}{0}
        \renewcommand{\theequation}{S\arabic{equation}}%
        \setcounter{table}{0}
        \renewcommand{\thetable}{S\arabic{table}}%
        \setcounter{figure}{0}
        \renewcommand{\thefigure}{S\arabic{figure}}%
     }

\definecolor{myred}{RGB}{214,26,70}
\definecolor{myreddark}{RGB}{76,8,38}
\definecolor{myblue}{RGB}{35,106,185}
\definecolor{mybluedark}{RGB}{19,56,99}
\definecolor{mybluebright}{RGB}{225,236,249}

\def\te{{\rm e}}

\def\bd{{\bf d}}
\def\bk{{\bf k}}
\def\bp{{\bf p}}
\def\bq{{\bf q}}
\def\br{{\bf r}}

\def\calH{\mathcal{H}}

\def\calK{\mathcal{K}}

\def\p{\hat {\psi}}

\def\T{{\mathcal T}}

\def\pa{\partial}
\def\nn{\nonumber}

\def\TB{ {\mathcal{T}_{\rm B} } }
\def\tB{{t_{\text{B}}}}
\def\aB{{a_{\text{B}}}}

\def\Re{{ \rm Re }}
\def\Im{{ \rm Im }}
\def\tp{{ \tilde{p} }}
\def\tq{{ \tilde{q} }}
\def\tk{{ \tilde{k} }}

\def\BEC{{ \rm BEC }}

\def\bpsi{{\pmb \psi}}
\def\bu{{\bf u}}

\begin{document}
\title{Superfluid flow of polaron polaritons above Landau's critical velocity}
\author{K.\ Knakkergaard Nielsen}
\affiliation{Department of Physics and Astronomy, Aarhus University, Ny Munkegade, 8000 Aarhus C, Denmark}
\author{A.\ Camacho-Guardian} 
\affiliation{Department of Physics and Astronomy, Aarhus University, Ny Munkegade, 8000 Aarhus C, Denmark}
\author{G.\ M.\ Bruun} 
\affiliation{Department of Physics and Astronomy, Aarhus University, Ny Munkegade, 8000 Aarhus C, Denmark}
\affiliation{Shenzhen Institute for Quantum Science and Engineering and Department of Physics, Southern University of Science and Technology, Shenzhen 518055, China}
\author{T.\ Pohl}
\affiliation{Department of Physics and Astronomy, Aarhus University, Ny Munkegade, 8000 Aarhus C, Denmark}
\date{\today}

\begin{abstract}
We develop a theory for the interaction of light with superfluid optical media, describing the motion of quantum impurities that are created and dragged through the liquid by propagating photons. It is well known that a mobile impurity suffers dissipation due to phonon emission as soon as it moves faster than the speed of sound in the superfluid -- Landau's critical velocity. Surprisingly we find that in the present hybrid light-matter setting, polaritonic impurities can be protected against environmental decoherence and be allowed to propagate well above the Landau velocity without jeopardizing the superfluid response of the medium. 
\end{abstract}

\maketitle

When an object moves through a superfluid it can do so without friction as long as it is slower than a certain critical velocity. In his seminal work \cite{Landau1941}, Landau obtained this bound by arguing that a moving impurity can generate excitations only when it exceeds the speed of sound in the superfluid. In this case, the object emits Cherenkov radiation which decelerates its motion. Being a hallmark of superfluidity this effect and the associated Landau velocity have since been investigated in diverse systems, from liquid helium \cite{Allum1977,Brauer2013,Bradley2016} and exciton-polariton fluids in semiconductor microcavities \cite{Amo2009}, to ultracold atomic quantum gases \cite{Raman1999}. 

An atomic impurity inside an ultracold gas of bosonic atoms \cite{Devreese2009,Schmidt2013,Casteels2014,Li2014,Levinsen2015,Ardila2015,Christensen2015,Schmidt2018,Ichmoukhamedov2019} provides an ideally suited and well controllable platform to study such behavior, as demonstrated in recent experiments \cite{Arlt2016,Jin2016,killian2018,Zwierlein2019}. These measurements revealed the emergence of a polaron quasiparticle in close analogy to its solid-state counterpart, introduced more than 80 years ago \cite{Landau1933,Frohlich1952} to understand how electrons interact with lattice vibrations of the surrounding crystal. The underlying Fr\"ohlich model \cite{Frohlich1952} has since found applications to various problems. For example, light-matter interactions originate from the optical generation of excitations in the material, whereby the coupling \cite{Frohlich1952} between such excitons and phonons can lead to dissipation and explains some important optical properties of semiconductors \cite{Toyozawa58}. The realization of strong light-matter coupling in such systems has enabled broad explorations of collective phenomena \cite{Amo2009,Adams2010,Carusotto2013,Jager2016,Leonard2017,Munoz-Matutano2019,Bao2019} and future applications \cite{Ballarini2013,Jariwala2014,Barachati2018,Schneider2018,Scuri2018,Back2018,Walther2018,gu2019} of exciton-polaritons. However, their coupling to phonons and ensuing damping of polarons remains a major limiting factor for coherence and quantum effects in such systems. 

\begin{figure}[th!]
\begin{center}
\includegraphics[width=0.95\columnwidth]{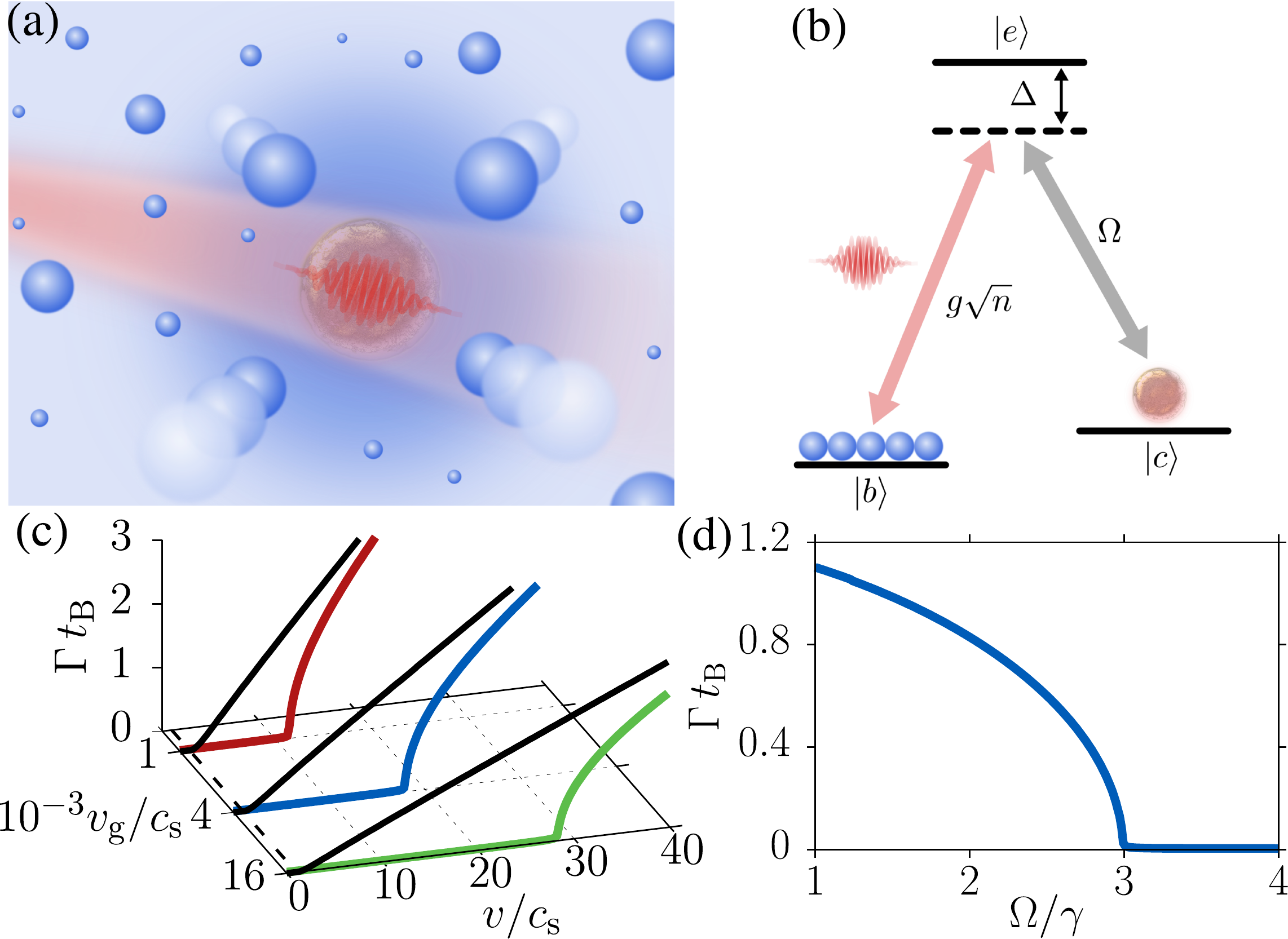}
\caption{(a) Illustration of a propagating photon generating a dark-state polaron-polariton via impurity interactions. (b) The incident photon can form a dark-state polariton by coupling the atomic ground state $\ket{b}$ to an excited state $\ket{e}$ with a detuning $\Delta$ and a coupling strength $g\sqrt{n}$, determined by the atomic density $n$. The state $\ket{e}$ decays radiatively with rate $\gamma$ and is coupled to a stable impurity state $\ket{c}$ via a classical control field with Rabi frequency $\Omega$. Panels (c) and (d) show the decay rate $\Gamma$ of the formed polaron-polariton in units of $\tB = \xi / \sqrt{2}c_{\rm s}$, determined by the coherence length, $\xi$, and the speed of sound, $c_{\rm s}$, of the superfluid. (c) $\Gamma$ as a function of the impurity speed $v$ and the polariton group velocity $v_{\rm g}$, varied through the density, $n \simeq 2 \cdot10^{14}\:{\rm (red)}, 0.8\cdot10^{14}\:{\rm (blue)}, 0.3\cdot10^{14}{\rm cm}^{-3}\:{\rm (green)}$, for $\Omega / \gamma = 2$ and $\Delta / \gamma = -200$. The damping of the  bare polaron is shown by the black lines. (d) $\Gamma$ as a function of $\Omega$ for $n \simeq 0.8\cdot10^{14}{\rm cm}^{-3}$ and $\Delta / \Omega = -300$, revealing the emergence of a critical field $\Omega / \gamma \sim 3$. All calculations are performed for the ${\rm D}_1$ transition of ultracold $^{23}$Na atoms and an impurity scattering length of $a = 500 a_0$, in units of the Bohr radius $a_0$.}
\label{fig.fig1}
\end{center}
\end{figure}

Here, we address this issue by developing a theory for the non-equilibrium dynamics of polaritons in a quantum many-body system under the formation of Fr\"ohlich polarons [see Fig. \ref{fig.fig1}(a)]. Considering the three-level scheme illustrated in Fig. \ref{fig.fig1}(b), we demonstrate the emergence of polaron-polariton quasiparticles that can vastly exceed the traditional Landau critical velocity of the medium without suffering phonon-induced decoherence [see Fig. \ref{fig.fig1}(c)]. This effect, in turn, permits to stabilize and protect an otherwise decaying polaron against phonon-induced decoherence via a vanishingly small photon-component of the formed polariton [see Fig. \ref{fig.fig1}(d)]. The discovery of such unusual behavior sheds new light on the optical properties of quantum many-body systems and may open up new routes for controlling and mitigating phonon-induced decoherence in light-matter interfaces.

More specifically, we consider a superfluid medium consisting of a weakly interacting atomic Bose-Einstein condensate (BEC), whereby an incident photon may transfer an atom to a different internal quantum state, which then acts as an impurity. Its interaction with the surrounding superfluid generates phonons, which screen the impurity to form a polaronic quasiparticle. To avoid dissipation from radiative decay of the excited state $\ket{e}$, one can apply an additional control field and couple two stable atomic states, the state $\ket{b}$ comprising the BEC and the state $\ket{c}$ being the impurity state, via a two-photon transition as shown in Fig. \ref{fig.fig1}(b). On two-photon resonance, the depicted three-level scheme realizes electromagnetically induced transparency (EIT), which affords strong light-matter coupling at virtually vanishing photon losses \cite{Fleischhauer2002} due to the formation of so-called dark-state polaritons \cite{Fleischhauer2000} that propagate with a greatly reduced group velocity, $v_{\rm g}$, as low as a few m/s \cite{Hau1999}.  At such low group velocities, the dark-state polariton is primarily composed of the impurity excitation with a very low photon fraction less than $10^{-6}$\cite{Fleischhauer2000}.  

Taken separately, these scenarios thus yield two stable quasiparticles: a photon-dressed impurity and a phonon-dressed impurity, which remains stable as long as its velocity is below the Landau velocity, i.e. the speed of sound in the superfluid. Consequently, one would expect that the combined quasiparticle destroys superfluidity \cite{Fleischhauer2016} as soon as $v_{\rm g}$ exceeds Landau's critical velocity. Surprisingly, this is not the case. First, it turns out that it is not the group velocity which determines the viscosity of its environment, but the total recoil momentum exerted on the impurity state by the two applied light fields. The resulting impurity velocity $v$, is widely tunable via the angle between the two laser fields and can differ vastly from $v_{\rm g}$. Second, we show that \emph{both} of these velocities of the moving impurity can greatly exceed Landau's critical velocity without destroying the superfluid response of the quantum liquid [see Fig. \ref{fig.fig1}(c)].

In order to understand these findings, let us consider a BEC of atoms with a mass $m$, a density $n$, and three internal states $\ket{b}$, $\ket{e}$ and $\ket{c}$, which are coupled by the propagating quantum light field and a classical control laser as indicated in Fig. \ref{fig.fig1}(b). We focus on weak collisional interactions that are short-ranged and can be parametrized by a scattering length $a_{\rm B}$ for the condensate atoms in the ground state $\ket{b}$ and a scattering length $a$ quantifying the interaction between the impurity atoms in the $\ket{c}$-state and the condensate. The underlying Hamiltonian $\hat{H} = \hat{H}_0 + \hat{H}_{\rm int} + \hat{H}_{\rm al}$ \cite{SM} can be conveniently split into three parts. Here, 
\begin{gather}
\!\hat{H}_0 = \sum_{\bp} \!\varepsilon^{\alpha}_\bp \hat{\alpha}_{\bf p}^\dagger \hat{\alpha}_{\bf p} \!+ \sum_\bk \!\left[ \varepsilon^{\rm e}_\bk \hat{e}^\dagger_\bk \hat{e}_\bk + \varepsilon^{\rm c}_\bk \hat{c}^\dagger_\bk \hat{c}_\bk + \omega_\bk \hat{\beta}^\dagger_\bk \hat{\beta}_\bk \!\right]
\end{gather}
describes one-body energies of the incident photons and the atoms in the atomic states $\ket{e}$, and $\ket{c}$, which are respectively created by the operators $\hat{\alpha}_{\bf p}^\dagger$ for a given momentum $\bp$, and $\hat{e}_{\bf k}^\dagger$, $\hat{c}_{\bf k}^\dagger$ with a given momentum $\bf k$. We consider a narrow-band incoming photon field, propagating along the $z$-axis with momenta $\bp$ that are tightly centered around the carrier momentum $\bp_0=p_0{\bf e}_z$. This defines a rotating frame in which the photon energy is $\varepsilon^{\alpha}_\bp = c (p - p_0)$, with the speed of light $c$. The complex energy $\varepsilon^{\rm e}_\bk = k^2 / 2m + \Delta - i \gamma$ of excited-state atoms contains the one-photon detuning $\Delta$ and decay rate $\gamma$, while the energy $\varepsilon^{\rm c}_\bk = k^2 / 2m + \delta$ of the impurity state is set by the two-photon detuning $\delta$. Excitations of the weakly interacting condensate are Bogoliubov modes, created by $\hat{\beta}^\dagger_\bk = u_\bk \hat{b}_\bk^\dagger + v_\bk \hat{b}_{-\bk}$ at momenta $\bk$ with energy $\omega_\bk$, whereby $b^\dagger_\bk$ creates an atom in the atomic ground state $\ket{b}$ and $u_\bk$, $v_\bk$ are the corresponding BEC coherence factors \cite{Stringari}. The light-matter interaction, 
\begin{align}
\hat{H}_{\rm al} =& \Omega \sum_{\bk} \hat{c}^\dagger_{\bk - \bk_{\rm cl}} \hat{e}_\bk + \frac{g}{\sqrt{V}} \sum_{\bk, \bp} \hat{b}^\dagger_{\bk} \hat{\alpha}^\dagger_{\bp} \hat{e}_{\bp + \bk} +{\rm h.c.},
\label{eq.H_al_single_mode}
\end{align}
describes the coupling to the classical control field with wave vector $\bk_{\rm cl}$ and Rabi frequency $\Omega$, as well as the single-photon interaction with a coupling strength $g$ within the rotating wave approximation. While the sum over $\bp$ is restricted to momenta for the incident photons, the photonic vacuum has been integrated out \cite{SM} yielding the decay rate $\gamma$ of the excited state included in $\varepsilon^{\rm e}_\bk$ above. In the absence of atomic interactions and at the two-photon resonance $\delta = 0$, the dynamics governed by $\hat{H}_0 + \hat{H}_{\rm al}$ shows that incoming photons are converted to dark-state polaritons $\hat{d}_\bp=\cos\theta \hat{\alpha}_\bp - \sin \theta \hat{c}_{\bp - \bk_{\rm cl}}$ that propagate the medium without losses at a velocity $v_{\rm g}=\cos^2 \theta\, c$, determined by $\tan\theta=g\sqrt{n}/\Omega$ \cite{Fleischhauer2000}. The typical case of large single-photon Rabi frequencies $g\sqrt{n}\gg\Omega$ \cite{Hau1999}, thus effectively yields an impurity $\hat{c}_{\bp - \bk_{\rm cl}}\approx -\hat{d}_\bp$ that has a form stable propagation through the condensate with an ultraslow velocity $v_{\rm g} \ll c$.

The interaction between the impurity and the superfluid can be described by the Fr{\"o}hlich Hamiltonian \cite{Frohlich1952}
\begin{equation}
\hat{H}_{\rm int} = \frac{\sqrt{n}\T}{\sqrt{V}} \sum_{\bq, \bk} (u_\bk - v_\bk)\hat{c}^\dagger_{\bq - \bk}\hat{c}_\bq\left(\hat{\beta}^\dagger_\bk + \hat{\beta}_{-\bk}\right),
\label{eq.H_bc}
\end{equation}
which serves as a paradigmatic model for a range of solid-state systems \cite{Frohlich1952,Alexandrov2010} and applies to polarons in BECs with weak interactions \cite{Christensen2015}. Physically, Eq. \eqref{eq.H_bc} describes momentum-changing impurity collisions that generate Bogoliubov excitations with an underlying scattering matrix $\T = 4\pi a / m$. These collisions can profoundly alter the idealized scenario of dissipation-free polariton motion. \\

\begin{figure}[t!]
\begin{center}
\includegraphics[width=\columnwidth]{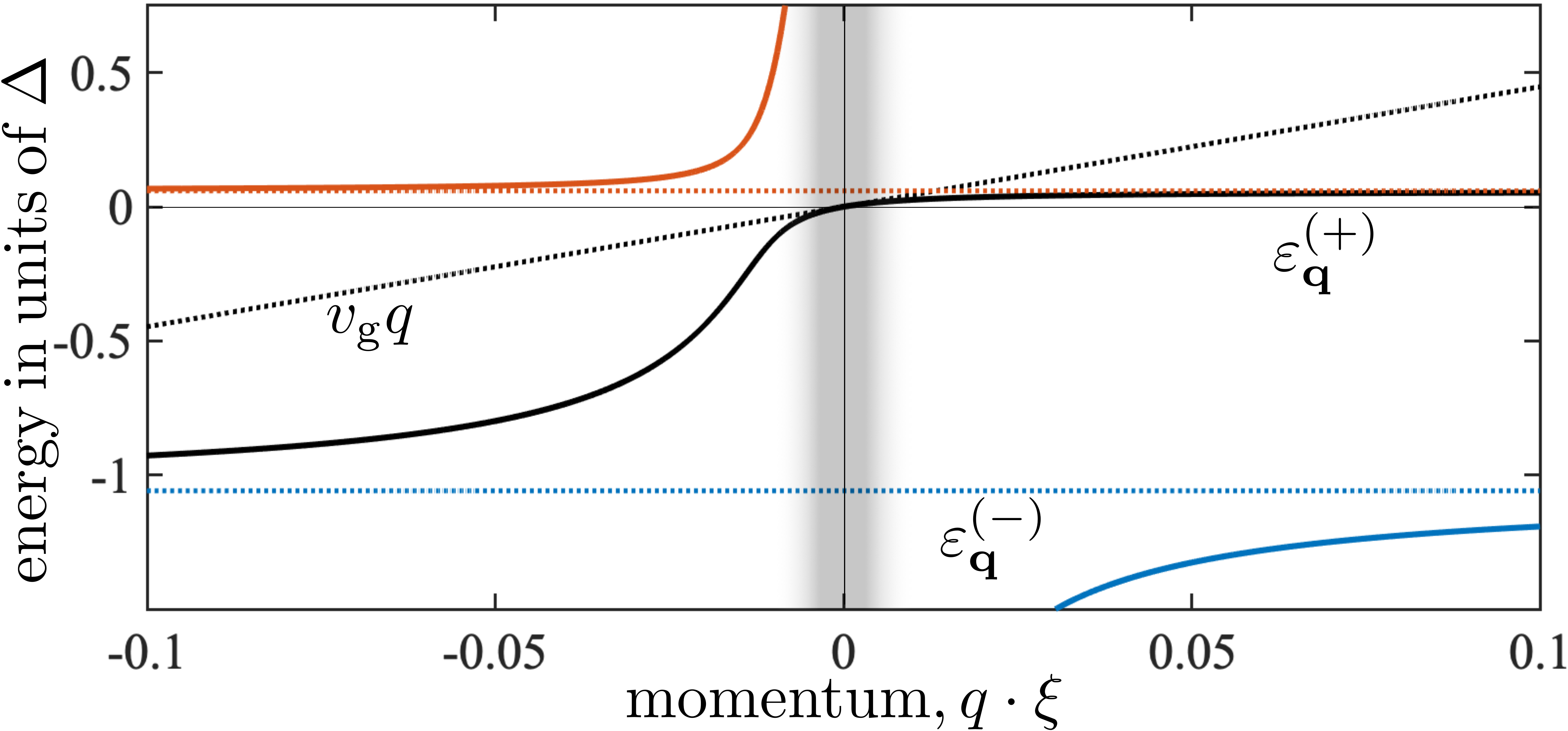}
\caption{Polariton dispersion curves in the absence of atomic interactions for $\Delta = -200\gamma$, $\Omega = 2\gamma$, and $n = 0.5 \cdot 10^{14} {\rm cm}^{-3}$. (a) Incoming photons generate dark-state polaritons (black solid line) with an approximate linear dispersion, $\varepsilon_\bp \simeq v_{\rm g}(p - p_0) + (\bp - \bk_{\rm cl})^2 / 2m$, around $p-p_0\approx0$ (black dotted line) that facilitates low-loss form stable photon propagation with the slow-light group velocity $v_{\rm g}$. 
Atomic collisions with the surrounding condensate cause a typical momentum change of $\Delta p\sim1 / \xi$ well outside this EIT regime, indicated by the vertical grey bar. The dark state is thereby broken apart by any atomic collision event, and scatters into the photon-free hybridized states $\ket{\pm}$, with the indicated energies $\varepsilon^{(\pm)}_{\bf p}$, shown by the orange and blue dashed lines in panel (a) and (b). This characteristic scattering process leads to the ansatz Eq.~\eqref{eq.state_ansatz} for the polaron-polariton. As illustrated in panel (b), the energy of the state $\ket{-}$ is typically so far removed that it does not contribute significantly to the emerging polaron-polariton quasiparticle and its self-energy, Eq.~\eqref{eq.self_energy}. Panel (c) shows the same dispersion curves on an expanded momentum scale, revealing the quadratic contribution from the atomic kinetic energy and the light shift induced by the classical control field.}
\label{fig.fig2}
\end{center}
\end{figure}

To characterize the resulting many-body dynamics, we use an ansatz
\begin{gather}
\ket{\Psi_\bp(t)} 	= \left[ A^{(0)}_\bp(t) \hat{\alpha}^\dagger_{\bp} + E^{(0)}_\bp(t) \hat{e}^\dagger_{\bp} + C^{(0)}_\bp(t) \hat{c}^\dagger_{\bp - \bk_{\rm cl}}\right]\ket{\BEC} \nn \\
					+ \sum_{\bk}\! \left[E^{(1)}_{\bp, \bk}(t) \hat{e}^{\dagger}_{\bp - \bk} + C^{(1)}_{\bp, \bk}(t) \hat{c}^{\dagger}_{\bp -\bk_{\rm cl} - \bk} \right]\! \hat{\beta}^{\dagger}_{\bk} \ket{\BEC},
\label{eq.state_ansatz}
\end{gather}
for the time-dependent wave function, which is truncated at the single phonon level to leading order in the impurity interaction. Here $\ket{\BEC}$ denotes the initial state of the Bose-Einstein condensate composed entirely of $\ket{b}$-state atoms. The first line describes the bare photon-driven impurity dynamics that yields the loss-less propagation of the dark-state polariton amplitude $D_\bp = \bra{\BEC}d_{\bp}\ket{\Psi_\bp(t)} = \cos\theta A^{(0)}_\bp - \sin\theta C^{(0)}_\bp$ discussed above. Collisions between the impurity and the surrounding atoms, however, perturb this polariton state and excite the superfluid as described by the Fr\"ohlich term in Eq. \eqref{eq.H_bc} and captured by the second line in Eq. \eqref{eq.state_ansatz}. The characteristic momentum change associated with such collisions is given by the inverse coherence length $1 / \xi = \sqrt{8\pi n\aB}$ of the condensate, which for a large single-photon detuning, $|\Delta| \gg \gamma$, lies far outside the EIT regime. Consequently, almost all impurity collisions, apart from negligible scattering events around $|\bp - \bk| \simeq p$ \cite{SM}, lead to a break up of the low-energy dark-state polariton and populate the hybridized modes $\ket{\pm}$ of the two laser-coupled $\ket{e}$- and $\ket{c}$-states with energies $\varepsilon^{(\pm)}_{\bf p} = [\varepsilon^{\rm e}_{\bf p} + \varepsilon^{\rm c}_{{\bf p} - {\bf k}{\rm cl}} \pm (4\Omega^2 + (\varepsilon^{\rm e}_{{\bf p}} - \varepsilon^{\rm c}_{{\bf p} - {\bf k}{\rm cl}})^2)^{1/2}]/2$ as indicated in Fig. \ref{fig.fig2}(a) and (b). This implies a prompt photon loss and is reflected in the omission of the photon component in the second line of Eq. \eqref{eq.state_ansatz}. It is this interaction-induced modification of the polariton character and associated dispersions that causes the unusual propagation phenomena found in this work. 

By using this ansatz in the many-body Schr{\"o}dinger equation $i\pa_t\ket{\Psi_\bp} = \hat{H}\ket{\Psi_\bp}$ we obtain a set of coupled equations for the five state amplitudes in Eq. \eqref{eq.state_ansatz}. Upon solving the evolution equations for $E^{(1)}$ and $C^{(1)}$ and substituting the result into the equations for the zero phonon amplitudes, we derive a closed equation \cite{SM}  
\begin{equation}
i\pa_t D_\bp(t) = [\varepsilon_\bp + \Sigma_{\bp} - \tilde{\Sigma}_\bp(t) ]D_\bp(t)
\label{eq.dark_state_eq_of_motion}
\end{equation}
that describes the open quantum dynamics of the dark-state polariton due to its interaction with the surrounding superfluid. Here, $\varepsilon_\bp = v_{\rm g} (p - p_0) + \sin^2\theta (\bp - \bk_{\rm cl})^2 / 2m$ is the dispersion of the non-interacting dark state polariton around $p_0$ [see Fig. \ref{fig.fig2}(a)]. The second term accounts for the kinetic energy of the atoms and is normally discarded when describing slow-light propagation \cite{Fleischhauer2000, Fleischhauer2002}. Here, however, it plays a crucial role in capturing the physics of atomic interactions. The time-dependent complex energy $\tilde{\Sigma}_\bp(t)$ \cite{SM} captures the non-equilibrium dynamics driven by the atomic interactions following the creation of the ideal dark state polariton at time $t = 0$. The vanishing of $\tilde{\Sigma}_\bp(t)$ at longer times then signals the establishment of a new quasiparticle -- the polaron-polariton. Its self-energy 
\begin{align}
\Sigma_\bp =&\int \frac{{\rm d}^3 k}{(2\pi)^3}\left[\frac{\left(g^{(+)}_{\bp, \bk}\right)^2}{\varepsilon_{\bp} - \varepsilon^{(+)}_{\bp - \bk} - \omega_\bk} + \frac{\left(g^{(-)}_{\bp, \bk}\right)^2}{\varepsilon_{\bp} - \varepsilon^{(-)}_{\bp - \bk} - \omega_\bk}\right. \nn \\
&\left. + \sin^2\theta \cdot n\T^2 \frac{m}{k^2}\right]
\label{eq.self_energy}
\end{align}
describes the effects of interactions on the quasiparticle dispersion and has a simple physical interpretation. First note that the classical control field hybridizes the $\ket{e}$- and $\ket{c}$-states of the atoms and generates new dressed states $\ket{\pm}$ with energies $\varepsilon^{(\pm)}_{\bp}$, as outlined above and indicated in Fig. \ref{fig.fig2}. Equation \eqref{eq.self_energy} therefore describes the virtual scattering of the impurity into these hybridized modes $\ket{\pm}$ upon the generation of phonon excitations with an energy $\omega_\bk$. The associated coupling elements \cite{SM}
\begin{align}
g^{(+)}_{{\bf p}, {\bf k}} &= \sin\theta \left[ u^{\rm ec}_{{\bf p} - {\bf k}} \sqrt{n}{\mathcal T}(u_{\bf k} - v_{\bf k}) + w^{\rm ec}_{{\bf p} - {\bf k}}\frac{v_{\bf k} \Omega}{\sqrt{n}}\right], \nonumber \\
g^{(-)}_{{\bf p}, {\bf k}} &= \sin\theta \left[ w^{\rm ec}_{{\bf p} - {\bf k}} \sqrt{n}{\mathcal T}(u_{\bf k} - v_{\bf k}) - u^{\rm ec}_{{\bf p} - {\bf k}}\frac{v_{\bf k} \Omega}{\sqrt{n}} \right]
\label{eq.effective_couplings}
\end{align} 
are determined by the form of the hybridized states, described by $u^{\rm ec}_{{\bf q}} = (\varepsilon^{(+)}_{{\bf q}} - \varepsilon^{\rm e}_{{\bf q}} ) / [( \varepsilon^{(+)}_{{\bf q}} - \varepsilon^{\rm e}_{{\bf q}})^2 + \Omega^2]^{1/2}$ and $w^{\rm ec}_{{\bf q}} = \Omega / [( \varepsilon^{(+)}_{{\bf q}} - \varepsilon^{\rm e}_{{\bf q}})^2 + \Omega^2]^{1/2}$, whereby $g^{(-)}_{\bp}$ vanishes as $g^{(-)}_{\bp}\sim\Omega$ with a decreasing control field. Eventually, Eq. \eqref{eq.self_energy} approaches the known second order polaron energy \cite{Casteels2014} in the zero-field limit in which the dark-state polariton coincides with the bare impurity. The obtained equation of motion \eqref{eq.dark_state_eq_of_motion} has a simple solution $D_{\bp}(t) = D_{\bp}(0) \te^{-iE_{\bp}\, t - \Gamma_{\bp}\,t}\te^{i\int_0^{t}{\rm d}\tau \tilde{\Sigma}_\bp(\tau)}$. Starting from an initially non-interacting dark-state polariton, $D_{\bp}(0)$, this solution describes the initial quasiparticle formation, as determined by $\tilde{\Sigma}_\bp(t)$, and the subsequent evolution of the formed polaron-polariton, governed by its energy $E_{\bp} = \varepsilon_\bp + \Re\Sigma_{\bp}$ and steady-state damping rate $\Gamma_\bp = -\Im\Sigma_{\bp}$. In the more familiar case of a bare polaron ($\Omega=0$), the impurity suffers a finite damping rate, $\Gamma_\bp$, if it moves faster than the Landau critical velocity, given by the condensate's speed of sound $c_{\rm s}=\sqrt{4\pi a_{\rm B}n}/m$. The kinetic energy is then sufficient to generate phonon excitations with a low-energy dispersion $\omega_\bk \simeq c_{\rm s} k$ and cause dissipation in the form of Cherenkov radiation \cite{Landau1941}. However, the damping rate of our dark-state polaron-polariton, shown in Fig. \ref{fig.fig1}(c), suggests profoundly different behavior than this paradigmatic scenario for the breakdown of superfluidity.

We observe that the group velocity, $v_{\rm g}$, which governs the speed with which the impurity excitation traverses the medium, has virtually no bearing on the damping of the polaron and can exceed $c_{\rm s}$ by several orders of magnitude. In fact, it turns out that it is not the velocity $v_{\rm g}$ of the polaritonic quasiparticle that determines the superfluid response of the medium, but the velocity of the laser-excited impurity atom. This velocity, ${\bf v}=({\bf p}-{\bf k}_{\rm cl})/m$, can be widely tuned via the propagation angle between the incident control laser and the probe photons with wave vectors $\bk_{\rm cl}$ and $\bp \simeq \bp_0$, respectively. 

Yet, even this velocity can exceed the speed of sound of the condensate by more than an order of magnitude without jeopardizing its superfluid response, as shown in Fig. \ref{fig.fig1}(c). To understand this behavior, we consider the off-resonant limit, $\Omega / |\Delta| \ll 1$, in which the $\ket{-}$-state is far removed in energy as shown in Fig. \ref{fig.fig2}(b), whereby the term involving $g^{(-)}_{\bp,\bk}$ in Eq. \eqref{eq.self_energy} can be neglected. As a result, the denominator of the first term in Eq. \eqref{eq.self_energy} dictates the energy balance
\begin{equation}
\frac{(\bp - \bk_{\rm cl})^2}{2m} = \frac{(\bp - \bk_{\rm cl} - \bk)^2}{2m}+ \omega_\bk - \frac{\Omega^2}{\Delta},
\label{eq.energy_balance}
\end{equation}
for the scattering of a polariton with energy $\varepsilon_{\bp}$ into a different momentum state with $\varepsilon_{\bp-\bk}^{(+)} \simeq (\bp - \bk_{\rm cl} - \bk)^2/2m - \Omega^2/\Delta$ while emitting a phonon with an energy $\omega_\bk$ via collisions between the impurity and its surrounding atoms. To obtain Eq. \eqref{eq.energy_balance}, we set $\varepsilon_{\bp}\simeq(\bp - \bk_{\rm cl})^2/2m$, since $\sin\theta \simeq 1$, and because the photon momentum $p$ is well within the EIT window such that $v_{\rm g}|p-p_0|\ll\Omega^2/|\Delta|$ [see Fig. \ref{fig.fig2}(a)]. Without the light field ($\Omega=0$), Eq. \eqref{eq.energy_balance} permits phonon emission only for impurity velocities $v=|\bp - \bk_{\rm cl}|/m\geq c_{\rm s}$ above the familiar Landau critical velocity $v_{\rm c}=c_{\rm s}$. In contrast, the presence of the light field renders the impurity collisions inelastic by introducing an additional energy cost $-\Omega^2/\Delta$ associated with the collisional break up of the dark-state polariton into the laser-dressed $\ket{+}$-state impurity as indicated in Fig. \ref{fig.fig2}(c). For a positive single-photon detuning, $\Delta>0$, the resulting endothermic character of the impurity collisions promotes phonon emission regardless of the impurity speed, corresponding to a vanishing critical velocity, $v_{\rm c}=0$. 

\begin{figure}[t!]
\begin{center}
\includegraphics[width=\columnwidth]{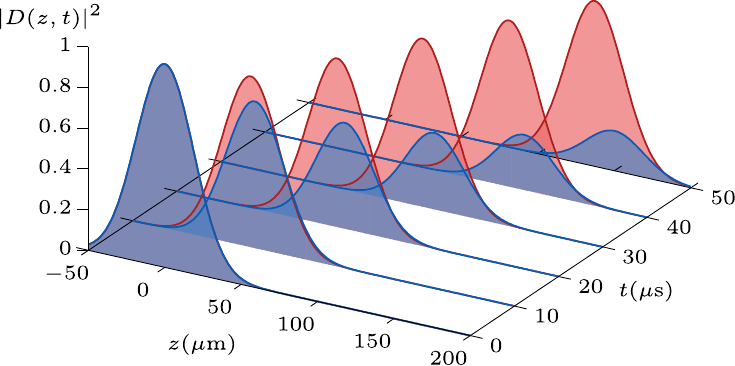}
\caption{Pulse propagation through a condensate of $^{23}$Na atoms with a density of $n = 2.6 \cdot 10^{14} {\rm cm}^{-3}$ and an impurity scattering length of $a = 0.1\xi$. The dynamics of a bare impurity wave packet (blue lines) suffers strong damping due to the supersonic motion of the formed polaron with an initial velocity of $3{\rm m/s}\gg c_{\rm s}$. In contrast, the red lines show the asymptotically undamped motion of a polaron-polariton with $\Delta = -200\gamma$ and an identical initial group velocity of $v_{\rm g}=3{\rm m/s}$, corresponding to a near-unity impurity fraction of $1-v_{\rm g}/c=0.99999999$. Polaron formation eventually leads to a slight lowering of the group velocity \cite{SM}.}
\label{fig.fig3}
\end{center}
\end{figure}\

A negative detuning, $\Delta<0$, on the other hand, introduces an additional energy cost for impurity collisions and thereby increases the critical velocity. Upon increasing the light shift $\Omega^2/\Delta$, this effect can indeed cause a substantial enhancement and increase the critical velocity by more than an order of magnitude under typical conditions of ultracold atom experiments \cite{Hau1999}. At the same time, this effect enables the quantum optical stabilization of otherwise decaying polaron quasiparticles. Indeed, Fig. \ref{fig.fig1}(d) reveals the emergence of a critical behavior with respect to the control field amplitude and demonstrates the efficient protection of the polaron against the otherwise inevitable emission of Cherenkov radiation above a critical control field $\Omega_{\rm c} \simeq v\sqrt{-m\Delta/4}$ \cite{SM}. 

This optical stabilization of the Bose polaron against phonon emission can be probed directly by measuring the transmission of slow-light polaritons through an ultracold gas of Bose condensed atoms. The propagation dynamics through the gas is conveniently visualized by Fourier transforming the obtained solution, $D_\bp(t)$, into real space. Figure \ref{fig.fig3} compares the resulting pulse evolution for a bare Bose polaron and a dark-state polaron-polariton, moving at initially identical velocities through a $^{23}$Na condensate with experimentally accessible densities and laser parameters. The Bose polaron undergoes rapid decoherence due to the steady emission of Cherenkov radiation \cite{SM}, while the amplitude of the dark-state polaron-polariton settles at the quasiparticle residue \cite{Knakkergaard2019} and remains otherwise protected from decoherence, eventually propagating at a lowered group velocity $v_{\rm g}+\partial_{p}{\rm Re}\Sigma_{\bf p}|_{{\bf p}_0}$. \\

The demonstrated ability to stabilize mobile polaritons in a dissipative environment thus provides an intriguing outlook for realizing coherent optical interfaces and makes it possible to explore and control the combined formation of polaritonic and polaronic quasiparticle states at greatly reduced losses and decoherence. Not only does this combination yield an attractive platform for exploring impurity physics \cite{Fleischhauer2016}, and suggest novel optical probes of quantum many-body dynamics \cite{Camacho2019}, but also promises new functionalities for light-matter interfaces and optical devices \cite{Sidler2016,Tan2019}. In the present context, ensuing applications include the generation of few-photon nonlinearities via induced polaron interactions in atomic superfluids \cite{Guardian2018,Guardian2018b}, which may even be controlled and enhanced via resonant phonon-exchange processes. Moreover, as outlined above, the underlying interaction Hamiltonian \eqref{eq.H_bc} is of considerably greater applicability describing for example the coupling between excitons and phonons in semiconductors \cite{Alexandrov2010}, which often presents a limitation to the coherence of light-matter interactions in such systems \cite{Toyozawa58}. The EIT-enabled stabilization against phonon-induced dissipation, described in this work, therefore suggests a promising approach to alleviating this obstacle. These combined perspectives motivate future investigations into the strong-coupling regime as well as a wider range of environmental interactions and photon interfaces for exploiting correlated quantum dynamics and exploring quantum nonlinear optics in strongly interacting many-body systems. 

\tocless\acknowledgments
The authors thank Luis Pe\~{n}a Ardila, Michael Fleischhauer and Eugene Demler for helpful discussions. This work has been supported by the Villum Foundation and the Independent Research Fund Denmark - Natural Sciences via Grant No. DFF - 8021-00233B, by the EU through the H2020-FETOPEN Grant No. 800942640378 (ErBeStA), by the DFG through the SPP1929, by the Carlsberg Foundation through the Semper Ardens Research Project QCooL, and by the DNRF through a Niels Bohr Professorship to TP. 

\bibliographystyle{apsrev4-1}
\let\oldaddcontentsline\addcontentsline
\renewcommand{\addcontentsline}[3]{}
\bibliography{ref_polaron_polariton}
\let\addcontentsline\oldaddcontentsline

\cleardoublepage
\onecolumngrid
\appendix

\beginsupplement
\tocless\section{\large Supplemental Material: Superfluid flow of polaron-polaritons above Landau's critical velocity}

\tableofcontents

\section{The Hamiltonian}

The quantum light field is described by the Hamiltonian
\begin{equation}
\hat{H}_{\rm l} = \sum_{\bk\lambda} ck \cdot \hat{\alpha}^\dagger_{\bk\lambda} \hat{\alpha}_{\bk\lambda}, 
\label{eq.H_l_initial}
\end{equation}
where $ck$ is the energy of a photon at momentum $\bk$ and polarization $\lambda$ created by $\hat{\alpha}^\dagger_{\bk\lambda}$. The atom-light coupling consists of a classical control field and a quantum field. For the former, we use $\mathbf{E} = \bm{\epsilon} \cdot E_0 \cos(\bk_{\rm cl} \cdot \br - c k_{\rm cl} t)$, with a (classical) wave vector $\bk_{\rm cl}$ and polarization vector $\bm{\epsilon}$. In the dipole approximation the Hamiltonian in first quantization is $-e \br_e \cdot \mathbf{E}$, with $\br_e$ the position vector of the electron relative to the atomic nucleus. With the Rabi frequency $\Omega = -e E_0 \bra{e} \br \cdot \bm{\epsilon}\ket{c} / 2$ we can then write the classical control field in second quantization,
\begin{align}
\hat{H}_{\rm al}^{(1)} &= 2 \Omega \int {\rm d}^3 r \left[ \p^\dagger_{\rm c}(\br)\p_{\rm e}(\br) \cos(\bk_{\rm cl} \cdot \br - c k_{\rm cl} t) + {\rm h.c.} \right] \simeq \Omega \int {\rm d}^3 r \left[ \p^\dagger_{\rm c}(\br)\p_{\rm e}(\br) {\rm e}^{-i(\bk_{\rm cl} \cdot \br - c k_{\rm cl} t)} + {\rm h.c.} \right] \nn \\
			 &= \Omega \sum_{\bk} \left[ \hat{c}^\dagger_{\bk - \bk{\rm cl}} \hat{e}_\bk {\rm e}^{i ck_{\rm cl}t} + {\rm h.c.} \right] = \Omega \sum_{\bk} \left[ \hat{\tilde{c}}^\dagger_{\bk - \bk{\rm cl}}(t) \hat{\tilde{e}}_\bk(t) + {\rm h.c.} \right],
			 \label{eq.H_al_classical_field}
\end{align}
where $\p^\dagger_{\rm a}(\br)$ creates an atom in state $\ket{a}$ at position $\br$. In the second equality we make the usual rotating wave approximation. In the second line we first transform to momentum space using $\p_{\rm a}(\br) = \sum_{\bk} {\rm e}^{i \bk \cdot \br} \hat{a}_{\bk} / \sqrt{V}$ for $a = e, c$, with $V$ the volume of the gas. Hence, $\hat{a}^\dagger_\bk$ creates an atom in state $\ket{a}$ at momentum $\bk$. We finally describe the Hamiltonian in the frame rotating with the light fields, using $\hat{\tilde{e}}_{\bk} = \hat{e}_\bk \cdot {\rm e}^{icp_0t}$ and $\hat{\tilde{c}}_\bk = \hat{c}_\bk \cdot {\rm e}^{ic(k_{\rm cl} - p_0)t}$. Here $c p_0$ is the carrier frequency of the  quantum light field, which we now turn to. We describe the coupling to the quantum light field in terms of a quantized electric field 
\begin{equation}
\hat{\mathbf{E}}(\br) = \frac{1}{\sqrt{V}}\sum_{\bk\lambda} \sqrt{\frac{ck}{2\epsilon_0}}\left[ \bm{\epsilon}_{\bk\lambda} \hat{\alpha}_{\bk\lambda} {\rm e}^{i\bk\cdot \br} + {\rm h.c.} \right],
\end{equation}
with $\epsilon_0$ the vacuum permittivity. The field is transverse: $\bk \cdot \bm{\epsilon}_{\bk \lambda} = 0$. We then get
\begin{align}
\hat{H}_{\rm al}^{(2)} &= \int {\rm d}^3 r \left[\p^\dagger_{\rm b}(\br) \p_{\rm e}(\br) \mathbf{d}_{\rm be} \cdot \hat{\mathbf{E}}(\br) + {\rm h.c.} \right] \simeq \frac{1}{\sqrt{V}} \sum_{\bk, \bq,\lambda} g_{\bk - \bq\lambda} \left[\hat{b}^\dagger_{\bq} \hat{\tilde{e}}_{\bk}(t) \hat{\tilde{\alpha}}^\dagger_{\bk - \bq\lambda}(t) + {\rm h.c.} \right],
\end{align}
with the electric dipole moment $\mathbf{d}_{\rm be} = -e\bra{b}\br \ket{e}$. In turn $g_{\bk\lambda} = \sqrt{ck / 2\epsilon_0} \; \bm{\epsilon}_{\bk \lambda} \cdot \mathbf{d}_{\rm be}$. We again make the rotating wave approximation and write the fields in the rotating frame, with $\hat{\tilde{\alpha}}_{\bk\lambda} = \hat{\alpha}_{\bk\lambda} {\rm e}^{icp_0t}$ a temporally slowly varying field when $k \simeq p_0$. We can describe the Hamiltonian in terms of time-independent fields, if we further adjust the energies of the photons and the atomic excited and impurity states. Hence, we write 
\begin{equation}
\hat{H}_0 = \sum_{\bk} \left[ \varepsilon^{\alpha}_{\bk}\sum_{\lambda}\hat{\alpha}^\dagger_{\bk\lambda} \hat{\alpha}_{\bk\lambda} + \left(\xi_\bk + \Delta \right)\hat{e}^\dagger_{\bk}\hat{e}_{\bk} + \left(\xi_\bk + \tilde{\delta} \right)\hat{c}^\dagger_{\bk}\hat{c}_{\bk} + \omega_{\bk} \hat{\beta}_{\bk}^\dagger \hat{\beta}_{\bk}\right].
\label{eq.H_0}
\end{equation}
The first term describes the photons, where we shift the energy by $cp_0$ in the rotating frame, letting $\varepsilon^{\alpha}_{\bk} = c(k - p_0)$. The second term describes the excited state $\ket{e}$ with the one-photon detuning $\Delta = \varepsilon^{\rm e}_0 - cp_0$, $\varepsilon^{\rm e}_0$ being the bare energy of the state. Also, $\xi_\bk = k^2 / 2m$ is the kinetic energy. The third term describes the impurity state with the two-photon detuning $\tilde{\delta} = \varepsilon^{\rm c}_0 + c(p_0 - k_{\rm cl})$, $\varepsilon^{\rm c}_0$ being the bare energy of the state. We here dropped the $\sim$'s for simplicity. Finally, the fourth term is the usual expression for the BEC Hamiltonian with $\hat{\beta}^\dagger_\bk = u_\bk \hat{b}_\bk^\dagger + v_\bk \hat{b}_{-\bk}$ creating a Bogoliubov mode at momentum $\bk$ and energy $\omega_\bk = [\xi_\bk (\xi_\bk + 2n\TB)]^{1/2}$. $u_\bk, v_\bk = ((\xi_\bk + n\TB) / \omega_\bk \pm 1)^{1/2} / \sqrt{2}$ are the BEC coherence factors, $n$ is the density of the condensate, and $\TB = 4\pi a_{\rm B}/m$ the zero energy scattering matrix for the $\ket{b}$ atoms. With the rotating frame in place, we may write for the atom-light coupling 
\begin{equation}
\hat{H}_{\rm al} = \Omega \sum_{\bk} \left[ \hat{c}^\dagger_{\bk - \bk{\rm cl}} \hat{e}_\bk + {\rm h.c.} \right] + \frac{1}{\sqrt{V}} \sum_{\bk, \bq,\lambda} g_{\bk\lambda} \left[\hat{b}^\dagger_{\bq} \hat{\alpha}^\dagger_{\bk\lambda} \hat{e}_{\bk + \bq}   + {\rm h.c.} \right].
\label{eq.H_al}
\end{equation}
Further, the impurity state, $c$, interacts with the ground state atoms, which at weak interactions can be described by the Fr{\"o}hlich interaction
\begin{equation}
\hat{H}_{\rm int} = n \T \sum_{\bk} \hat{c}^\dagger_\bk \hat{c}_\bk + \frac{\sqrt{n}\T}{\sqrt{V}} \sum_{\bk, \bq} (u_\bk - v_\bk) \hat{c}^\dagger_{\bq - \bk} \hat{c}_{\bq}\left(\hat{\beta}^\dagger_{\bk} + \hat{\beta}_{-\bk} \right), 
\label{eq.H_bc_initial}
\end{equation}
where $\T = 4\pi a / m$ is the zero energy scattering matrix for the $b$-$c$ interaction. The atomic $\ket{c}$-$\ket{c}$, $\ket{c}$-$\ket{e}$, and $\ket{e}$-$\ket{e}$ interactions are absent under the assumption that only a single quantum of light is propagating, i.e. at most a single atom is excited. Further, we will not consider any interaction between the ground and excited state, $\ket{b}$-$\ket{e}$. The elementary Hamiltonian of the system is thus $\hat{H} = \hat{H}_0 + \hat{H}_{\rm al} + \hat{H}_{\rm int}$. 

\begin{figure}[th]
\begin{center}
\includegraphics[width=0.2\columnwidth]{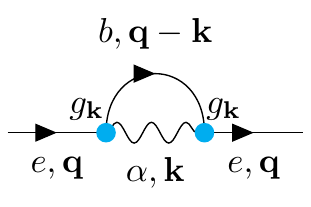}
\end{center}
\vspace{-0.5cm}
\caption{Feynman diagram for excited state to photonic vacuum coupling. This leads to a Lamb shift, incorporated in the energy $\varepsilon^{\rm e}_0$, and decay rate $\gamma$, see Eq. \eqref{eq.e_decay}.} 
\label{fig.Feynman_e_decay}
\end{figure}

We wish to simplify the Hamiltonian description to the incoming modes $\bp$ along the $z$-axis. This is accomplished by integrating out the photonic vacuum, i.e. all the photonic modes for $\bk \neq \bp$.  The Feynman diagram associated with this is shown in Fig. \ref{fig.Feynman_e_decay}, leading to the decay rate 
\begin{align}
\gamma_\bq &= \sum_{\lambda} \int \frac{{\rm d}^3 k}{(2\pi)^3} u_{\bq - \bk}^2 g_{\bk\lambda}^2 \delta(\xi_\bq + \Delta - \varepsilon^{\alpha}_\bk - \omega_{\bq - \bk}) \simeq \sum_{\lambda}\int \frac{{\rm d}^3 k}{(2\pi)^3} u_{\bq - \bk}^2g_{\bk\lambda}^2 \delta(\varepsilon^{\rm e}_0 - ck) \nn \\
&= \gamma + \sum_{\lambda}\int \frac{{\rm d}^3 k}{(2\pi)^3} v_{\bq - \bk}^2g_{\bk\lambda}^2 \delta(\varepsilon^{\rm e}_0 - ck).
\label{eq.e_decay}
\end{align}
In principle, we should omit the modes $\bp$ in this integration. However, because they lie along a single line the result is unaffected. Due to the huge slope of the photonic dispersion, the speed of light $c$, the atomic energies $\xi_\bq, \omega_{\bq - \bk}$ are completely negligible. In the second line we use $u_\bk^2 - v_\bk^2 = 1$. There is thus in principle a small correction to the bare (Wigner-Weisskopf) decay rate $\gamma = \sum_{\lambda}\int \frac{{\rm d}^3 k}{(2\pi)^3} g_{\bk\lambda}^2 \delta(\varepsilon^{\rm e}_0 - ck)$, scaling with the number of non-condensate $\ket{b}$-atoms. However, this has no bearing on our studies, and we will simply ignore it. 

For concreteness, we assume that the incoming photons are linearly polarized and that the direction of the electric dipole moment is fixed orthogonal to the propagation of the incoming photons. We can then set one of the polarizations, $\lambda_1$, to be parallel to the dipole moment. I.e. $\bm{\epsilon}_{\bp \lambda_1} \cdot \bd_{\rm be} = |\bd_{\rm be}|$ and $\bm{\epsilon}_{\bp \lambda_2} \cdot \bd_{\rm be} = 0$. This thus picks out a particular polarization, and defining $g = g_{\bp \lambda_1} = \sqrt{cp / 2\epsilon_0} \; |\mathbf{d}_{\rm be}|$ we may write an effective Hamiltonian describing only the incoming photonic modes, $\bp$, 
\begin{align}
\hat{H}_0 &= \sum_\bp \varepsilon^{\alpha}_{\bp} \hat{\alpha}^\dagger_\bp \hat{\alpha}_\bp + \sum_{\bk} \left[ \varepsilon^{\rm e}_\bk \hat{e}^\dagger_\bk \hat{e}_\bk + \varepsilon^{\rm c}_\bk \hat{c}^\dagger_\bk \hat{c}_\bk + \omega_\bk \hat{\beta}^\dagger_\bk \hat{\beta}_\bk \right], \nn \\
\hat{H}_{\rm al} &= \Omega \sum_{\bk} \left[ \hat{c}^\dagger_{\bk - \bk{\rm cl}} \hat{e}_\bk + {\rm h.c.} \right] + \sum_\bp \sqrt{n}g\left(\hat{\alpha}^\dagger_\bp \hat{e}_\bp + {\rm h.c.}\right) + \frac{g}{\sqrt{V}} \sum_{\bk,\bp} \left[(u_\bk\hat{\beta}^\dagger_\bk - v_\bk\hat{\beta}_{-\bk}) \hat{\alpha}^\dagger_{\bp} \hat{e}_{\bp + \bk}   + {\rm h.c.} \right], \nn \\
\hat{H}_{\rm int} &= \frac{\sqrt{n}\T}{\sqrt{V}} \sum_{\bk, \bq} (u_\bk - v_\bk) \hat{c}^\dagger_{\bq - \bk} \hat{c}_{\bq}\left(\hat{\beta}^\dagger_{\bk} + \hat{\beta}_{-\bk} \right).
\label{eq.H_eff}
\end{align}
Here we drop the now redundant polarization index $\lambda_1$ on $\alpha_{\bp \lambda_1}$. Also, $\varepsilon^{\rm e}_\bk = \xi_\bk + \Delta - i \gamma$ includes the decay rate of the excited state, and $\varepsilon^{\rm c}_\bk = \xi_\bk + \delta$ includes the mean field energy shift $n \T$ due to the impurity-boson interaction in the two-photon detuning $\delta = \tilde{\delta} + n\T$. 

\section{Deriving the equations of motion in the physical basis}
To accommodate for the atomic interactions and the quantum fluctuations in the BEC we use the state ansatz
\begin{align}
&\ket{\Psi_\bp}(t) = \left[ A^{(0)}_\bp(t) \hat{\alpha}^\dagger_{\bp} + E^{(0)}_\bp(t) \hat{e}^\dagger_{\bp} + C^{(0)}_\bp(t) \hat{c}^\dagger_{\bp - \bk{\rm cl}}\right]\ket{\BEC} + \sum_{\bk}\! \left[E^{(1)}_{\bp, \bk}(t) \hat{e}^{\dagger}_{\bp - \bk} + C^{(1)}_{\bp, \bk}(t) \hat{c}^{\dagger}_{\bp -\bk{\rm cl} - \bk} \right]\!\hat{\beta}^{\dagger}_{\bk}\ket{\BEC},
\label{eq.state_ansatz}
\end{align}
also given in the main text. Here the term describing an impurity plus a single phonon, $C^{(1)}$, is generated by the impurity-boson interaction $\hat{H}_{\rm int}$. This in turn is coupled to the $E^{(1)}$ term through the classical light field $\propto \Omega$. It is finally coupled to the photonic mode $A^{(0)}$ via terms in $\hat{H}_{\rm al}$ present due to quantum fluctuations $\propto v_\bk g$. The underlying assumption of this ansatz is that when the impurity scatters on the condensate atoms, it breaks apart the dark state and decouples the photonic mode from the atomic states. This is accurate when the typical scattering momentum is much larger than the largest change in momentum the dark state can suffer without breaking apart, $\Delta p_{\rm cr}$. In the weak coupling limit investigated here we have $k_{\rm scat} \simeq 1 / \xi$ with $\xi = 1 / \sqrt{8\pi n \aB}$ the BEC coherence length. On the other hand $\Delta p_{\rm cr}$ is determined by equating the energy at the edge of the EIT window $\Omega^2 / \sqrt{\Delta^2 + \gamma^2}$ with the dark state energy $v_{\rm g} \Delta p_{\rm cr}$, resulting in $\Delta p_{\rm cr} = n g^2 / (c\sqrt{\Delta^2 + \gamma^2})$. Thus, for the wave function ansatz to accurately describe the scattering, we need 
\begin{equation}
\frac{\Delta p_{\rm cr}}{k_{\rm scat}} = \frac{3\sqrt{\pi}}{2\sqrt{2}}\frac{1}{p^2}\sqrt{\frac{n}{\aB(1 + (\Delta / \gamma)^2)}} \ll 1, 
\label{eq.state_ansatz_accurate}
\end{equation}
where we use $g^2 = 3\pi c\gamma / p^2$. When this inequality is fulfilled the dark state breaks apart during an atomic scattering event as shown in Fig. 2 of the main text. The only caveat is when the scattering preserves the magnitude of the dark state momentum, i.e. $|\bp - \bk| \simeq p$. These events are however extremely rare and negligible as discussed at the end of this Supplemental Material. For atomic densities of $n = 10^{13} - 10^{15} {\rm cm}^{-3}$, optical transitions $\lambda = 2\pi / p = 400 - 800 {\rm nm}$, and typical atomic interactions in the condensate of $\aB = 100a_0$, $\Delta p_{\rm cr}/k_{\rm scat}$ is of order unity on the single photon resonance $\Delta = 0$. Therefore, the theory is restricted in validity to detunings much larger than the decay rate, $|\Delta| \gg \gamma$. Finally, terms with more than one phonon present will be higher order in the impurity-boson interaction, or macroscopically suppressed. E.g. there is in principle a term $\frac{1}{2}\sum_\bk A^{(2)}_{\bp, \bk}(t) \hat{\alpha}^{\dagger}_\bp \hat{\beta}^{\dagger}_\bk \hat{\beta}^{\dagger}_{-\bk}\ket{\BEC}$ coupling to the $E^{(1)}$ term through $g u_\bk$. However, this coupling turns out to be zero in the thermodynamic limit, where $N \to \infty, V \to \infty$, and we thus neglect it completely. We first solve the equations for the one phonon amplitudes $E^{(1)}$ and $C^{(1)}$ in terms of zero phonon amplitudes $A^{(0)}, E^{(0)}$ and $C^{(0)}$, and then plug these solutions back into the equations of motion for the zero phonon amplitudes. For convenience, we let $\bpsi^{(1)}_{\bp, \bk} = [E^{(1)}_{\bp, \bk}, \; C^{(1)}_{\bp, \bk}]$ and $\bpsi^{(0)}_{\bp} = [A^{(0)}_{\bp}, \; E^{(0)}_{\bp}, \; C^{(0)}_{\bp}]$. Using the Schr{\"o}dinger equation we then get
\begin{align}
i\pa_t \bpsi^{(1)}_{\bp, \bk}(t) &= \left[\omega_\bk + \calH_{\bp - \bk}\right] \bpsi^{(1)}_{\bp, \bk}(t) + \frac{1}{\sqrt{V}}G_\bk \bpsi^{(0)}_\bp(t), \nn \\
i\pa_t \bpsi^{(0)}_{\bp}(t) &= \calH_{\bp}^{(0)} \bpsi^{(0)}_{\bp, \bk}(t) + \frac{1}{\sqrt{V}} \sum_\bk G_\bk^\dagger \bpsi^{(1)}_{\bp,\bk}(t).
\label{eq.equations_of_motion_1}
\end{align}
Here, 
\begin{align}
\calH_{\bq} = \begin{bmatrix} \varepsilon^{\rm e}_{\bq} & \Omega \\ \Omega & \varepsilon^{\rm c}_{\bq - \bk{\rm cl}} \end{bmatrix}, \hspace{0.5cm} \calH^{(0)}_\bp = \begin{bmatrix} \varepsilon^{\alpha}_{\bp} & \sqrt{n}g & 0 \\ \sqrt{n}g & \varepsilon^{\rm e}_{\bp} & \Omega \\ 0 & \Omega & \varepsilon^{\rm c}_{\bp - \bk{\rm cl}} \end{bmatrix}, \hspace{0.5cm} G_\bk = \begin{bmatrix} -g v_\bk & 0 & 0 \\ 0 & 0 & \sqrt{n}\T (u_\bk - v_\bk) \end{bmatrix}.
\label{eq.Hamiltonian_1_0_and_coupling_matrix}
\end{align}
The first two describe the effective Hamiltonian of the scattered and unscattered states respectively, while $G_\bk$ describes the coupling matrix to the scattered states. The equation for these, $\bpsi^{(1)}_{\bp, \bk}$, in Eq. \eqref{eq.equations_of_motion_1} is formally solved to yield
\begin{equation}
\bpsi^{(1)}_{\bp, \bk}(t) = -\frac{i}{\sqrt{V}} \int_0^t {\rm d}\tau \; {\rm e}^{-i \omega_\bk (t - \tau)} {\rm e}^{-i\calH_{\bp - \bk} (t - \tau)} G_\bk \bpsi^{(0)}_{\bp}(\tau), 
\label{eq.formal_solution_H1_0} 
\end{equation}
using the initial condition $\bpsi^{(1)}_{\bp, \bk}(0) = {\bf 0}$, i.e. that there are no phonons initially. Reinserting this in the equation for $\bpsi^{(0)}_{\bp}$ in Eq. \eqref{eq.equations_of_motion_1} we get
\begin{equation}
i\pa_t \bpsi^{(0)}_\bp(t) = \calH^{(0)}_\bp\bpsi^{(0)}_{\bp}(t) - \frac{i}{V} \int_0^t {\rm d}\tau \sum_{\bk} {\rm e}^{-i\omega_\bk(t - \tau)} G_\bk^\dagger {\rm e}^{-i\calH_{\bp - \bk} (t - \tau)} G_\bk \bpsi^{(0)}_{\bp}(\tau).
\label{eq.equations_of_motion_2}
\end{equation}

We write out the explicit solution by finding eigenvectors and -values to $\calH_\bq$. The eigenvalues are $\varepsilon^{(\pm)}_\bq = \frac{1}{2}[\varepsilon^{\rm e}_{\bq} + \varepsilon^{\rm c}_{\bq - \bk{\rm cl}} \pm (4\Omega^2 + (\varepsilon^{\rm e}_{\bq} - \varepsilon^{\rm c}_{\bq - \bk{\rm cl}})^2)^{1/2}]$ describing hybridized modes $\ket{\pm}$ of the $\ket{e}$- and $\ket{c}$-states. The corresponding eigenvectors are
\begin{align}
\bu^{(+)}_{\bq} = \begin{bmatrix} w^{\rm ec}_{\bq} \\ u^{\rm ec}_{\bq} \end{bmatrix} = \frac{1}{\sqrt{( \varepsilon^{(+)}_{\bq} - \varepsilon^{\rm e}_{\bq})^2 + \Omega^2}}\begin{bmatrix} \Omega \\ \varepsilon^{(+)}_{\bq} - \varepsilon^{\rm e}_{\bq} \end{bmatrix}, \hspace{0.5cm} \bu^{(-)}_{\bq} = \begin{bmatrix} u^{\rm ec}_{\bq} \\ - w^{\rm ec}_{\bq} \end{bmatrix} = \frac{1}{\sqrt{( \varepsilon^{(+)}_{\bq} - \varepsilon^{\rm e}_{\bq})^2 + \Omega^2}}\begin{bmatrix} \varepsilon^{(+)}_{\bq} - \varepsilon^{\rm e}_{\bq} \\ - \Omega \end{bmatrix}.
\label{eq.eigenvectors_Hamiltonian_E1_C1}
\end{align}
The eigenmatrix $U_{\bq} = [\bu^{(+)}_{\bq} \; \bu^{(-)}_{\bq} ]$ is its own inverse, and so we get ${\rm e}^{-i \calH_{\bp - \bk} (t - \tau)} \bpsi^{(0)}_{\bp,\bk}(\tau) = {\rm e}^{-i \calH_{\bp - \bk} (t - \tau)}(U_{\bp - \bk})^2 \bpsi^{(0)}_{\bp,\bk}(\tau) = [\bu^{(+)}_{\bq} {\rm e}^{-i\varepsilon^{(+)}_{\bp - \bk}(t - \tau)} \; \bu^{(-)}_{\bq} {\rm e}^{-i\varepsilon^{(-)}_{\bp - \bk}(t - \tau)} ] U_{\bp - \bk} \bpsi^{(0)}_{\bp,\bk}(\tau)$. Performing the matrix multiplication, we get
\begin{align}
i\pa_t \begin{bmatrix} A^{(0)}_\bp \\[0.1cm] E^{(0)}_\bp \\[0.1cm] C^{(0)}_\bp \end{bmatrix} = \mathcal{H}^0_{\bp} \begin{bmatrix} A^{(0)}_\bp(t) \\[0.1cm] E^{(0)}_\bp(t) \\[0.1cm] C^{(0)}_\bp(t) \end{bmatrix} + i\int_0^{t} {\rm d}\tau \; \calK_{\bp}(t - \tau) \begin{bmatrix} A^{(0)}_\bp(\tau) \\[0.1cm] E^{(0)}_\bp(\tau) \\[0.1cm] C^{(0)}_\bp(\tau) \end{bmatrix}.
\label{eq.non_local_equations_of_motion}
\end{align}
Here,
\begin{align}
\mathcal{K}_{\bp}(t) &= \begin{bmatrix} \mathcal{K}_{\bp}^{\alpha\alpha}(t) & 0 & \mathcal{K}_{\bp}^{\rm \alpha c}(t) \\ 0 & 0 & 0 \\ \mathcal{K}_{\bp}^{\rm \alpha c}(t) & 0 &  \mathcal{K}_{\bp}^{\rm cc}(t)\end{bmatrix}, \hspace{0.5cm} \mathcal{K}_{\bp}^{\alpha\alpha}(t) = -g^2 \frac{1}{V}\sum_{\bk} v_\bk^2 \cdot {\rm e}^{-i\omega_\bk t} \left[(w^{\rm ec}_{\bp - \bk})^2 {\rm e}^{-i\varepsilon^{(+)}_{\bp - \bk} t} + (u^{\rm ec}_{\bp - \bk})^2 {\rm e}^{-i\varepsilon^{(-)}_{\bp - \bk} t}\right], \nn \\
\mathcal{K}_{\bp}^{\rm \alpha c}(t) &= \sqrt{n}g \T \frac{1}{V}\sum_{\bk} v_\bk (u_\bk - v_\bk) u^{\rm ec}_{\bp - \bk}w^{\rm ec}_{\bp - \bk} \cdot {\rm e}^{-i\omega_\bk t} \left[{\rm e}^{-i\varepsilon^{(+)}_{\bp - \bk} t} - {\rm e}^{-i\varepsilon^{(-)}_{\bp - \bk} t}\right], \nn \\
\mathcal{K}_{\bp}^{\rm cc}(t) &= - n\T^2 \frac{1}{V}\sum_{\bk} (u_\bk - v_\bk)^2 \cdot {\rm e}^{-i\omega_\bk t}  \left[ (u^{\rm ec}_{\bp - \bk})^2 {\rm e}^{-i\varepsilon^{(+)}_{\bp - \bk} t} + (w^{\rm ec}_{\bp - \bk})^2 {\rm e}^{-i\varepsilon^{(-)}_{\bp - \bk} t} \right].
\label{eq.K_effective}
\end{align}
We are now ready to transform to the polariton basis and make the equations of motion local in time. 

\section{Dark state equation of motion}
The polaritons are the eigenstates of the Hamiltonian,
\begin{align}
\calH^{(00)}_{\bp} = \begin{bmatrix} 0 & \sqrt{n}g & 0 \\ \sqrt{n}g & 0 & \Omega \\ 0 & \Omega & 0 \end{bmatrix}, \hspace{0.5cm} \ket{D_\bp} = \begin{bmatrix} \cos \theta \\ 0 \\ -\sin\theta \end{bmatrix}, \hspace{0.5cm} \ket{B^{(\pm)}_\bp} = \frac{1}{\sqrt{2}}\begin{bmatrix} \sin \theta \\ \pm 1 \\ \cos\theta \end{bmatrix},
\label{eq.H00}
\end{align}
with the eigenvectors given on the right, using $\tan\theta = \Omega / \sqrt{n}g$. The eigenvalues of these are $0$ for the dark state $\ket{D_\bp}$ and $\pm \sqrt{n g^2 + \Omega^2}$ for the two bright states $\ket{B^{(\pm)}_\bp}$. We thus define $W_{\bp} = \left[\ket{B^{(+)}_{\bp}} \; \ket{B^{(-)}_{\bp}} \; \ket{D_{\bp}}\right]$ and let
\begin{equation}
\begin{bmatrix} A^{(0)}_\bp \\[0.1cm] E^{(0)}_\bp \\[0.1cm] C^{(0)}_\bp \end{bmatrix} = W_{\bp} \begin{bmatrix} B^{(+)}_\bp \\[0.1cm] B^{(-)}_\bp \\[0.1cm] D_\bp \end{bmatrix},
\end{equation}
defining the polariton amplitudes $B^{(\pm)}_\bp$ and $D_\bp$. The equations of motion in Eq. \eqref{eq.non_local_equations_of_motion} transformed to the polariton basis is thus
\begin{align}
i\pa_t \begin{bmatrix} B^{(+)}_{\bp}(t) \\[0.1cm] B^{(-)}_{\bp}(t) \\[0.1cm] D_{\bp}(t) \end{bmatrix} = \bar{\calH}^0_\bp \begin{bmatrix} B^{(+)}_{\bp}(t) \\[0.1cm] B^{(-)}_{\bp}(t) \\[0.1cm] D_{\bp}(t) \end{bmatrix} + i \int_0^{t} {\rm d}\tau \; \bar{\calK}_{\bp}(t - \tau) \begin{bmatrix} B^{(+)}_{\bp}(\tau) \\[0.1cm] B^{(-)}_{\bp}(\tau) \\[0.1cm] D_{\bp}(\tau) \end{bmatrix}, 
\label{eq.non_local_equations_of_motion_polariton_basis}
\end{align}
with
\begin{align}
\!\! \bar{\calH}^{(0)}_\bp = W^\dagger_\bp \calH^{(0)}_\bp W_\bp = \begin{bmatrix} 
\varepsilon^{\rm B (+)}_\bp & \frac{\sin^2\theta\varepsilon^{\alpha}_{\bp} - \varepsilon^{\rm e}_{\bp} + \cos^2\theta \varepsilon^{\rm c}_{\bp - \bk{\rm cl}}}{2} 
& \frac{\cos \theta \sin\theta}{\sqrt{2}}(\varepsilon^{\alpha}_\bp -\varepsilon^{\rm c}_{\bp - \bk{\rm cl}}) 
\\ \frac{\sin^2\theta\varepsilon^{\alpha}_{\bp} - \varepsilon^{\rm e}_{\bp} + \cos^2\theta \varepsilon^{\rm c}_{\bp - \bk{\rm cl}}}{2} 
&\varepsilon^{\rm B (-)}_\bp&  \frac{\cos \theta \sin\theta}{\sqrt{2}}(\varepsilon^{\alpha}_\bp -\varepsilon^{\rm c}_{\bp - \bk{\rm cl}}) 
\\ \frac{\cos \theta \sin\theta}{\sqrt{2}}(\varepsilon^{\alpha}_\bp -\varepsilon^{\rm c}_{\bp - \bk{\rm cl}}) 
&  \frac{\cos \theta \sin\theta}{\sqrt{2}}(\varepsilon^{\alpha}_\bp -\varepsilon^{\rm c}_{\bp - \bk{\rm cl}}) 
& \varepsilon_\bp \end{bmatrix} ,
\label{eq.H0_polariton_basis}
\end{align}
defining the energies $\varepsilon^{\rm B(\pm)}_\bp = \pm (n g^2 + \Omega^2)^{1/2} +  (\sin^2\theta\varepsilon^{\alpha}_{\bp} + \varepsilon^{\rm e}_{\bp} + \cos^2\theta \varepsilon^{\rm c}_{\bp - \bk{\rm cl}}) / 2$, and $\varepsilon_\bp = \cos^2\theta \varepsilon^{\alpha}_\bp \!+\! \sin^2\theta\varepsilon^{\rm c}_{\bp - \bk{\rm cl}}$ for the bright and dark states respectively. Finally, 
\begin{align}
\bar{\calK}_\bp(t) &= W^\dagger_\bp \calK_\bp(t) W_\bp = \begin{bmatrix} \calK^{\rm B}_{\bp}(t) & \calK^{\rm B}_{\bp}(t) & \calK^{\rm BD}_{\bp}(t) \\[0.1cm] \calK^{\rm B}_{\bp}(t) & \calK^{\rm B}_{\bp}(t) & \calK^{\rm BD}_{\bp}(t) \\[0.1cm] \calK^{\rm BD}_{\bp}(t) & \calK^{\rm BD}_{\bp}(t) & \calK^{\rm D}_{\bp}(t) \end{bmatrix}, \\
\calK^{\rm B}_{\bp}(t) &= \frac{1}{2}\left(\sin^2\theta \cdot \calK^{\alpha\alpha}_\bp(t) + 2 \cos\theta \sin\theta \cdot \calK_{\bp}^{\rm \alpha c}(t) + \cos^2\theta \cdot \calK_{\bp}^{\rm cc}(t) \right), \nn \\
\calK^{\rm BD}_{\bp}(t) &= \frac{1}{\sqrt{2}}\left((\cos^2\theta - \sin^2\theta) \cdot \calK_{\bp}^{\rm \alpha c}(t) + \cos\theta \sin\theta \cdot (\calK_{\bp}^{\alpha\alpha}(t) - \calK_{\bp}^{\rm cc}(t))\right), \nn \\
\calK^{\rm D}_{\bp}(t) &=  \sin^2\theta \cdot \calK_{\bp}^{\rm cc}(t) - 2 \cos\theta \sin\theta\cdot \calK_{\bp}^{\rm \alpha c}(t) + \cos^2\theta \cdot \calK_{\bp}^{\alpha\alpha}(t).
\label{eq.polariton_non_local_coefficients}
\end{align}
We could keep all terms and propagate all three amplitudes, $B^{(+)}_\bp, B^{(-)}_\bp$ and $D_\bp$. However, because the bright states are so far removed in energy, by $\pm (n g^2 + \Omega^2)^{1/2}$, the dark and bright states effectively decouple. We therefore completely ignore the bright states, and rewrite $\calK^{\rm D}$ according to
\begin{align}
\calK^{\rm D}_{\bp}(t) &= \sin^2\theta \left[\calK_{\bp}^{\rm cc}(t) - \frac{2}{\tan\theta} \calK_{\bp}^{\rm \alpha c}(t) + \frac{1}{\tan^2\theta} \calK_{\bp}^{\alpha\alpha}(t)\right] = \sin^2\theta \left[\calK_{\bp}^{\rm cc}(t) - 2\frac{\Omega}{\sqrt{n}g} \calK_{\bp}^{\rm \alpha c}(t) + \left(\frac{\Omega}{\sqrt{n}g}\right)^2 \calK_{\bp}^{\alpha\alpha}(t)\right] \nn \\
&= -\frac{1}{V}\sum_{\bk}{\rm e}^{-i\omega_\bk t} \left[\left(g^{(+)}_{\bp, \bk}\right)^2{\rm e}^{-i\varepsilon^{(+)}_{\bp - \bk} t} + \left(g^{(-)}_{\bp, \bk} \right)^2{\rm e}^{-i\varepsilon^{(-)}_{\bp - \bk} t} \right],
\end{align}
with the effective couplings 
\begin{align}
g^{(+)}_{\bp, \bk} \!= \sin\theta \left[ u^{\rm ec}_{\bp - \bk} \sqrt{n}\T(u_\bk - v_\bk) + w^{\rm ec}_{\bp - \bk}\frac{v_\bk \Omega}{\sqrt{n}}\right], \hspace{0.5cm} g^{(-)}_{\bp, \bk} \!= \sin\theta \left[ w^{\rm ec}_{\bp - \bk} \sqrt{n}\T(u_\bk - v_\bk) - u^{\rm ec}_{\bp - \bk}\frac{v_\bk \Omega}{\sqrt{n}} \right].
\label{eq.effective_couplings}
\end{align}
We are now ready to compute the time-local equation of motion for the dark state. Perturbatively consistent we set $D_{\bp}(\tau) = {\rm e}^{i\varepsilon_{\bp}(t - \tau)}D_{\bp}(t)$ in the temporal integral in Eq. \eqref{eq.non_local_equations_of_motion_polariton_basis}, and get
\begin{align}
i\pa_t D_\bp = \left(\varepsilon_\bp + K_\bp(t)\right)D_\bp(t),
\label{eq.dark_state_equation_of_motion}
\end{align}
with $K_\bp(t) = i \int_0^{t} {\rm d}\tau \; \calK^{\rm D}_{\bp}(t - \tau){\rm e}^{i\varepsilon_{\bp}(t - \tau)}$. Finally, we renormalize the impurity-boson interaction by adding $\sin^2\theta \cdot n\T^2 / V \cdot \sum_{\bk} m/k^2$ to $K_\bp$, making the equations fully consistent to second order in $\T$. Thus, in the above equation of motion $K_\bp(t)$ goes to $\Sigma_{\bp} - \tilde{\Sigma}_{\bp}(t)$, with the equilibrium self-energy 
\begin{align}
\Sigma_\bp = \int \frac{{\rm d}^3 k}{(2\pi)^3}\left[\frac{\left(g^{(+)}_{\bp, \bk}\right)^2}{\varepsilon_{\bp} - \varepsilon^{(+)}_{\bp - \bk} - \omega_\bk} + \frac{\left(g^{(-)}_{\bp, \bk}\right)^2}{\varepsilon_{\bp} - \varepsilon^{(-)}_{\bp - \bk} - \omega_\bk} + \sin^2\theta \cdot n\T^2 \frac{m}{k^2}\right],
\label{eq.self_energy}
\end{align}
also given in Eq. (6) of the main text and the time-dependent contribution
\begin{align}
\tilde{\Sigma}_\bp(t) = \int &\frac{{\rm d}^3 k}{(2\pi)^3}\left[ \left(g^{(+)}_{\bp, \bk}\right)^2\frac{{\rm e}^{i(\varepsilon_{\bp} - \varepsilon^{(+)}_{\bp - \bk} - \omega_\bk)t}}{\varepsilon_{\bp} - \varepsilon^{(+)}_{\bp - \bk} - \omega_\bk} + \left(g^{(-)}_{\bp, \bk}\right)^2\frac{{\rm e}^{i(\varepsilon_{\bp} - \varepsilon^{(-)}_{\bp - \bk} - \omega_\bk)t}}{\varepsilon_{\bp} - \varepsilon^{(-)}_{\bp - \bk} - \omega_\bk}\right].
\end{align}
We here replace the sum over momentum modes with integrals: $\sum_\bk / V \to \int {\rm d}^3 k / (2\pi)^3$. The effective couplings $g^{(\pm)}$ thus describe scattering into the hybridized $\ket{e}$-$\ket{c}$ states $\ket{\pm}$ through the generation of phonons. The dark state equation of motion, Eq. (5) in the main text, is thus obtained. \\

To clarify the interaction scalings we put $\tilde{\Sigma}$ on unitless form. We let the $z$-axis be in the direction of $\bp$. Writing explicitly the effective couplings $g^{(\pm)}_{\bp, \bk}$ then yields
\begin{align}
\tB \cdot \tilde{\Sigma}_\bp(t) = -\frac{2\sqrt{2}}{\pi^2} \frac{\aB}{\xi} \sin^2\theta \int_0^{2\pi}{\rm d}\varphi\int_{-1}^{+1}{\rm d}\cos\theta \int_0^{\infty}{\rm d}\tk \, \tk^2 &\left[ \left(u^{\rm ec}_{\bp - \bk}(u_\bk - v_\bk) \frac{a}{2\aB} + w^{\rm ec}_{\bp - \bk} v_{\bk}\tilde{\Omega}\right)^2 \frac{{\rm e}^{-i\Phi^{(+)}_{\bp, \bk}\tilde{t}}}{\Phi^{(+)}_{\bp, \bk}} \right. \nn \\ 
&\left.\!\! + \left(w^{\rm ec}_{\bp - \bk}(u_\bk - v_\bk) \frac{a}{2\aB} - u^{\rm ec}_{\bp - \bk} v_{\bk}\tilde{\Omega}\right)^2 \frac{{\rm e}^{-i\Phi^{(-)}_{\bp, \bk}\tilde{t}}}{\Phi^{(-)}_{\bp, \bk}}\right], 
\label{eq.Sigma_tilde_unitless}
\end{align}
with $\Phi^{(\pm)}_{\bp, \bk} = - (\varepsilon_{\bp} - \varepsilon^{(\pm)}_{\bp - \bk} - \omega_\bk)\tB$, $\tk = k\xi / \sqrt{2}$, $\tilde{t} = t / \tB$, and $\tilde{\Omega} = \Omega \tB$. The self-energy $\Sigma$ can be brought on a similar form. This shows that there are essentially three types of terms. The first scale as $a^2 / \aB\xi \cdot (u_\bk - v_\bk)^2$, the second as $a / \xi \cdot \tilde{\Omega} v_\bk(u_\bk - v_\bk)$ and the third as $\aB / \xi \cdot (\tilde{\Omega}v_\bk)^2$. 

\section{Critical velocity and Rabi frequency}
In this section we compute the critical velocity and Rabi frequency in the limit of $\Omega / |\Delta| \ll 1$ relevant for Figs. 1(c) and 1(d) of the main text. \\

In this limit the coupling to the $\ket{-}$ state vanishes, $g^{(-)} \to 0$. The critical behaviour is thus kinematically set by when the dark state can scatter into the $\ket{+}$ state in an energy conserving way, i.e. $\varepsilon_{\bp} = \Re [\varepsilon^{(+)}_{\bp - \bk} + \omega_{\bk} ]$ for some phonon momentum $\bk$, as evident from the self-energy $\Sigma_{\bp}$ (Eq. \eqref{eq.self_energy}). Expanding this equation to leading order in $\Omega / \Delta$, using $\varepsilon_\bp \simeq \varepsilon^{\rm c}_{\bp - \bk {\rm cl}}$ for $p = p_0$, and $\varepsilon^{\rm c}_{\bp - \bk{\rm cl}} = \xi_{\bp - \bk{\rm cl}} + \delta$ we must check when 
\begin{equation}
\xi_\bq = \xi_{\bq - \bk} - \frac{\Omega^2}{\Delta} + \omega_\bk
\label{eq.critical_v_1}
\end{equation}
can be solved as a function of $\bk$, with $\bq = \bp - \bk_{\rm cl}$. This equation also follows from a simple second order argument: to 2nd order in $\Omega$ the scattered impurity experiences a light shift $- \Omega^2/\Delta$, and thus alters the energy balance. If the scattered phonon is emitted in the forward direction of the impurity, the impurity kinetic energy $\xi_{\bq - \bk} = (\bq - \bk)^2 / 2m$ is lowered most significantly, and this is thus where we first get a solution. So we focus on $\bk \parallel \bq$. Writing Eq. \eqref{eq.critical_v_1} in units of $\tB$ we must then solve
\begin{equation}
0 = (\tq - \tk)^2 + \tk\sqrt{1 + \tk^2} - \tq^2 - \frac{\Omega^2}{\Delta} \tB = f(\tq, \tk) - \frac{\Omega^2}{\Delta} \tB. 
\label{eq.critical_v_2}
\end{equation}
For $\Delta > 0$ there always exists a solution to this equation, and therefore the critical velocity is 0 for positive detunings. For $\Delta < 0$ the light shift is always positive. Therefore, $f(\tq, \tk)$ must be negative for some interval of $\tk$ for a solution to exist. This only happens when $\tq = q \xi / \sqrt{2} > 1 / 2$ or equivalently when $v = q / m = |\bp - \bk_{\rm cl}| / m > c_{\rm s}$. This shows that the speed still needs to be larger than the speed of sound, as one might expect. However, because of the light shift, there may still not be a solution to Eq. \eqref{eq.critical_v_2}, leading to an increased critical velocity $v_{\rm c}$. We can compute this by finding the minimum of $f(\tp, \tk)$ and equating it to the light shift. While the general solution is rather involved, we can find a simple approximate solution for $v \gtrsim 2 c_{\rm s}$. Here we can approximate $\tk\sqrt{1 + \tk^2} \simeq \tk^2$ at the minimum. Taking the derivative, $\pa_{\tk} f(\tq, \tk) \simeq 4 (\tk - \tq / 2)$, then yields a minimum at $\tk \simeq \tq / 2$, and thus $\min_{\tk} f(\tq, \tk) \simeq f(\tq, \tq / 2) \simeq - \tq^2 / 2$. The critical momentum is thus $\tq_{\rm c} \simeq \sqrt{-2 \Omega^2 \tB / \Delta}$, yielding a critical velocity
\begin{equation}
v_{\rm c} = \frac{q_{\rm c}}{m} \simeq \frac{2\Omega}{\sqrt{-m\Delta}}, 
\label{eq.critical_v_final}
\end{equation}
using $\tq = q \xi / \sqrt{2}$ and $\tB = m\xi^2$.  
Hence, only if the impurity moves with a speed faster than $v_{\rm c}$ does it experience decay, as shown in Fig. 1(c). This gives an accurate result for $v_{\rm c} \gtrsim 2 c_{\rm s}$. Reversely, for a fixed velocity $v$ as in Fig. 1(d), one can increase the critical velocity by increasing $\Omega$. This thus exhibits a critical behaviour at 
\begin{equation}
\Omega_{\rm c} \simeq \frac{\sqrt{-m\Delta}v}{2}, 
\label{eq.critical_Rabi}
\end{equation}
\textit{below} which the dark state experiences decay, while above the decay rate becomes vanishingly small, as shown in Fig. 1(d). Again this is accurate for $v \gtrsim 2 c_{\rm s}$. The underlying reason for the vanishing decay rate is thus that the additional energy cost from the light field becomes too large at $\Omega_{\rm c}$ for the scattering to be allowed kinematically. 

\section{Propagation of the dark state}
\begin{figure}[t]
\begin{center}
\includegraphics[width=0.9\columnwidth]{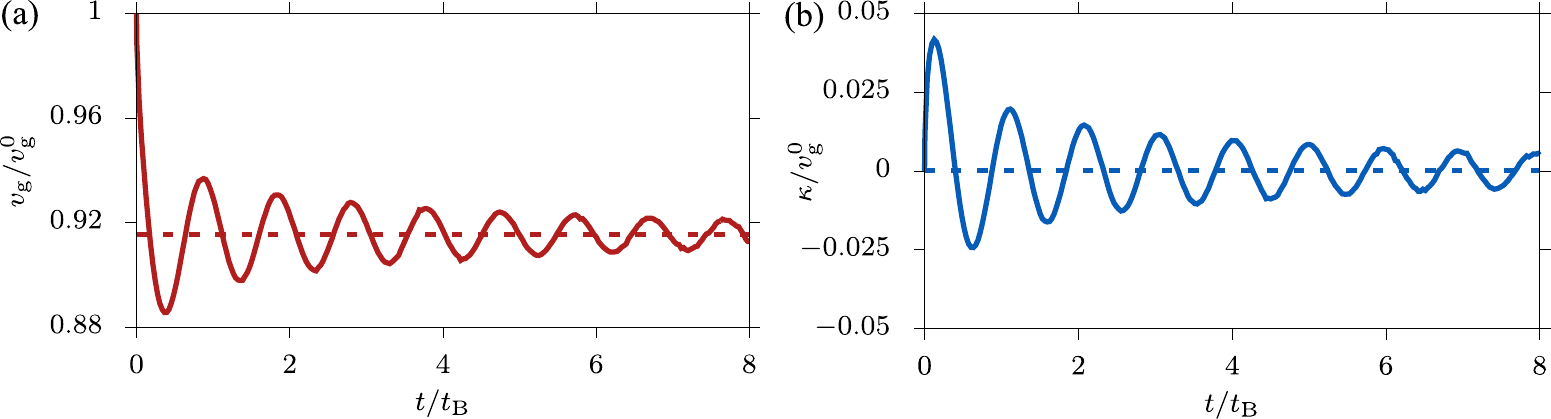}
\caption{Time-dependent group speed $v_{\rm g}$ (a) and damping coefficient $\kappa$ (b) in units of the group speed in the absence of atomic interactions: $v_{\rm g}^0 = \Omega^2 / (\Omega^2 + n g^2) \cdot c$. The group speed eventually settles to a value a few percent below $v_{\rm g}^0$ (dashed red in (a)), while the damping coefficient $\kappa$ essentially settles at $0$ (dashed blue in (b)).}
\label{fig.figS2}
\end{center}
\end{figure}

We derive an expression for the propagation of the dark state in real space. First however, we need an expression for the group speed. We thus define
\begin{equation}
v_{\rm g}(t) - i \kappa(t) = \pa_{p}\left.\left(\varepsilon_{\bp} + \Sigma_{\bp} - \tilde{\Sigma}_{\bp}(t) \right) \right|_{\bp = \bp_0}.
\label{eq.vq_kappa_definition}
\end{equation}
Since the right hand side is in general a complex number there is both a contribution to the group speed $v_{\rm g}$ and a damping coefficient $\kappa$, and due to the presence of the time-dependent rate coefficient $\tilde{\Sigma}$ these depend on time as well. Keeping only the dominant terms proportional to the group speed in the absence of atomic interactions $v_{\rm g}^0 = \Omega^2 / (\Omega^2 + n g^2) \cdot c$, with $c$ the speed of light, we obtain
\begin{equation}
v_{\rm g}(t) - i \kappa(t) = v_{\rm g}^0 \left[ 1 - it \cdot \tilde{\Sigma}_{\bp_0}(t) + i\int_0^{t}{\rm d}\tau \; \tilde{\Sigma}_{\bp_0}(\tau) \right]. 
\label{eq.vq_kappa_result}
\end{equation}
Let us now turn to the propagation. Suppose that we prepare a non-interacting dark state pulse $D(z, t = 0)$ at time $t = 0$. The evolution of this state can be studied by expanding in plane waves (along the propagation axis $z$)
\begin{equation}
D(z, t) = \int \frac{{\rm d}p}{2\pi}{\rm e}^{i p z} D_{\bp}(t) = \int \frac{{\rm d}p}{2\pi}{\rm e}^{i p z} D_{\bp}(0) {\rm e}^{-iE_{\bp} t - \Gamma_{\bp}t}{\rm e}^{i\int_0^{t}{\rm d}\tau \tilde{\Sigma}_\bp(\tau)}, 
\end{equation}
where we in the second equality use that $D_{\bp}(t) = D_{\bp}(0) {\rm e}^{-iE_{\bp}\, t - \Gamma_{\bp}\,t} {\rm e}^{i\int_0^{t}{\rm d}\tau \tilde{\Sigma}_\bp(\tau)}$ as described in the main text, with $E_{\bp} = \varepsilon_{\bp} + {\rm Re}\Sigma_{\bp}$ the dark state energy and $\Gamma_{\bp} = - {\rm Im}\Sigma_\bp$ the dark state decay rate. This simple solution is accurate, provided the pulse fits within the EIT window, i.e. $v_{\rm g} \sigma_p < \Omega^2 / \sqrt{\Delta^2 + \gamma^2}$, with $\sigma_p$ the momentum standard deviation. Using the definition of the group speed and damping coefficient in Eq. \eqref{eq.vq_kappa_definition} we get
\begin{align}
D(z, t) &\simeq {\rm e}^{i p_0 z} {\rm e}^{-iE_{\bp_0} t}{\rm e}^{-\Gamma_{\bp_0}t}{\rm e}^{i\int_0^t {\rm d}\tau \tilde{\Sigma}_{\bp_0}(\tau)} \int \frac{{\rm d}p}{2\pi}{\rm e}^{i(p - p_0) \left(z - \int_0^t {\rm d}\tau [ v_{\rm g}(\tau) - i\kappa(\tau)] \right) } D_{\bp}(0) \nn \\ 
&= {\rm e}^{i p_0 ( z - z'(t))} {\rm e}^{-iE_{\bp_0} t}{\rm e}^{-\Gamma_{\bp_0}t}{\rm e}^{i\int_0^t {\rm d}\tau \tilde{\Sigma}_{\bp_0}(\tau)} \int \frac{{\rm d}p}{2\pi}{\rm e}^{i p z'(t) } D_{\bp}(0) \nn \\
&= {\rm e}^{i p_0 \int_0^t {\rm d}\tau [ v_{\rm g}(\tau) - i\kappa(\tau)]} {\rm e}^{-iE_{\bp_0} t}{\rm e}^{-\Gamma_{\bp_0}t}{\rm e}^{i\int_0^t {\rm d}\tau \tilde{\Sigma}_{\bp_0}(\tau)} \cdot D(z'(t), 0), \nn
\end{align}
with $z'(t) = z - \int_0^t {\rm d}\tau [ v_{\rm g}(\tau) - i\kappa(\tau)]$. The probability distribution consequently becomes
\begin{equation}
|D(z, t)|^2 = {\rm e}^{+ 2 p_0 \int_0^t {\rm d}\tau \kappa(\tau) } {\rm e}^{-2\Gamma_{\bp_0}t}{\rm e}^{-2\int_0^t {\rm d}\tau \Im\tilde{\Sigma}_{\bp_0}(\tau)} |D(z'(t), 0)|^2. 
\label{eq.D_z_solution} 
\end{equation}
This concludes the present derivation, and describes motion of the pulse at a time-dependent group velocity $v_{\rm g}(t)$. In Fig. \ref{fig.figS2} we plot $v_{\rm g}$ and $\kappa$ as a function of time for the same parameters considered in Fig. 3 of the main text. The atomic interactions eventually leads to a slight lowering of the group speed, while the damping coefficient $\kappa$ remains vanishingly small. An analysis of the asymptotic dynamics shows that the oscillation frequency is exactly the light shift $\Omega^2 / |\Delta|$, while the dominant non-equilibrium contribution to $v_{\rm g}$ at long times vanishes as ${\rm e}^{- \Omega^2 / \Delta^2 \cdot \gamma t} / t$. For $\Delta < 0$, $\Omega / |\Delta| \ll 1$, and below the critical velocity the dark state decay rate $\Gamma_{\bp}$ becomes vanishingly small stabilizing the pulse as evident in Fig. 3, and the overall pulse is only reduced by the square of the dark state residue: $Z = {\rm e}^{-\int_0^\infty {\rm d}\tau \Im\tilde{\Sigma}_{\bp_0}(\tau)}$. 

\section{Impurity damping rate at large speeds}
In Fig. 3 we make a comparison of the dark state propagation with an impurity shot through the condensate at the group speed $v_{\rm g}$. Since the group speed is several orders of magnitude larger than the speed of sound, $v_{\rm g} \gg c_{\rm s}$, the perturbative result for the damping rate would here give a dramatic overestimate. Instead we use the ladder approximation as described in Ref. [7]. At zero temperature the ladder approximation yields the self-energy $\Sigma(\bp, \omega) = n \T(\bp, \omega)$ with the scattering matrix 
\begin{equation}
\T(\bp, \omega) = \frac{\T}{1 - \T \cdot \Pi(\bp, \omega)}, 
\label{eq.impurity_scattering_matrix_ladder_approximation}
\end{equation}
written in terms of the zero-energy impurity-boson scattering matrix $\T = 4\pi a / m$ and the pair propagator 
\begin{equation}
\Pi(\bp, \omega) = \int \frac{{\rm d}^3 k}{(2\pi)^3}\left[\frac{u_\bk^2}{\omega - \xi_{\bp - \bk} - \omega_\bk + i{\rm e}a} + \frac{m}{k^2}\right] \simeq -i \frac{m^{3/2}}{4\pi}\sqrt{\omega - \frac{p^2}{4m}},
\label{eq.impurity_pair_propagator}
\end{equation}
with ${\rm e}a = 0^+$ a positive infinitesimal. In the second equality we use that at the very high energy we are interested in, $\omega = \xi_\bp = p^2 / 2m = mv_{\rm g}^2 / 2$, only the large momenta contribute. Hence, we can safely approximate $u_\bk \simeq 1$ and $\omega_\bk \simeq \xi_\bk$. This makes the integral analytically solvable, yielding the vacuum pair propagator. The impurity damping rate then becomes
\begin{equation}
\Gamma_{\rm imp} = - \Im \left[\Sigma\left(\bp, \frac{p^2}{2m}\right)\right] \simeq - \Im\left[n\T(\bp, \omega)\right] = n \frac{2\pi  p a^2}{m} \frac{1}{1 + (p a / 2)^2} \simeq \frac{8\pi n}{m p}.
\label{eq.impurity_damping_rate} 
\end{equation}
In the last equality we use that $p a \simeq 25 \gg 1$ for the parameters used in Fig. 3 of the main text. In this figure we thus plot the \textit{retrieval} probability distribution of the impurity, using that the damping $\Gamma_{\rm imp}$ gives the scattering rate out of the $\bp$ momentum state. 

\section{Additional damping rate around $|\bp - \bk| \simeq p$} \label{sec.additional_damping}
\begin{figure}[t]
\begin{center}
\includegraphics[width=0.9\columnwidth]{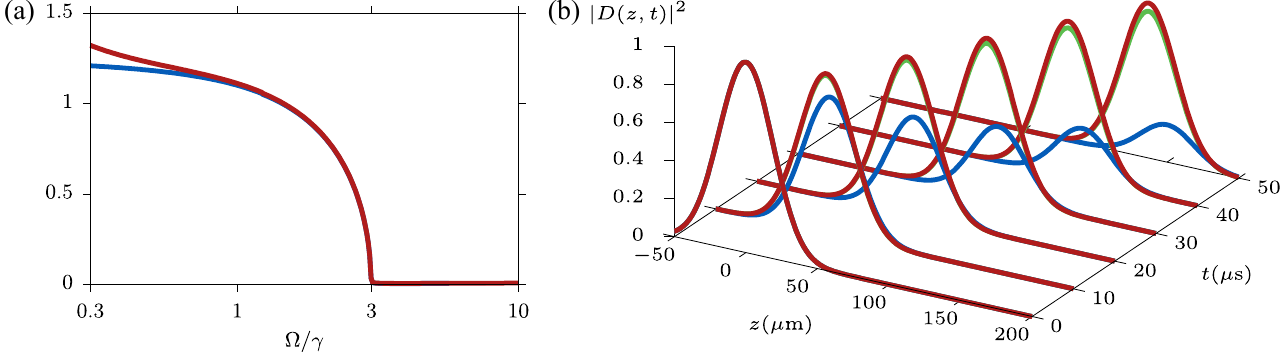}
\caption{(a) Scattering rate of the dark state. Plotted as a function of the Rabi frequency $\Omega$ of the classical control field. In red we show the corrected scattering rate $\Gamma + \Gamma_{\rm damp}$ including the dark state damping rate (Eq. \eqref{eq.Gamma_damp_approx}). In blue we show the dark state decay rate, $\Gamma$, also given in Fig. 1(d) of the main text. We only see deviations for very small $\Omega$ corresponding to $v_{\rm g} \sim c_{\rm s}$. We use the same parameters as in Fig. 1(d). (b) Propagation dynamics. The same pulse propagation as in Fig. 3 of the main text, with the bare impurity wave packet in blue, and the dark state wave packet in red. Here we include the additional dark state damping rate in Eq. \eqref{eq.Gamma_damp_approx} in the green lines. Importantly, we only see a very small correction to the red line. }
\label{fig.figS3}
\end{center}
\end{figure}

In the wave function ansatz \eqref{eq.state_ansatz}, equivalent to Eq. (4) of the main text, we have assumed that the dark state breaks apart in any atomic scattering event. However, there is a small probability for the dark state at momentum $\bp$ to scatter to other dark states. This happens when only the direction of the photonic momentum changes -- not the magnitude. I.e. when scattering a phonon with momentum $\bk$, the dark states survives when $|\bp - \bk| \simeq p$, defining a sphere of possible dark states. This leads to an additional damping rate $\Gamma_{\rm damp}$ of the dark state on top of the decay rate $\Gamma$ investigated in our present work. We calculate the damping rate from Fermi's golden rule, 
\begin{equation}
\Gamma_{\rm damp} = \sum_{\bk,\lambda} \left|\bra{f_{\bp, \bk, \lambda}} \hat{H}_{\rm int} \ket{i_{\bp, \lambda_i}}\right|^2 \pi \delta(\varepsilon_{\bp} - \varepsilon_{\bp - \bk} - \omega_{\bk}),
\label{eq.Gamma_damp_1}
\end{equation} 
with the initial state $\ket{i_{\bp,\lambda_i}} = \hat{d}^\dagger_{\bp,\lambda_i}\ket{\BEC}$ and the final states $\ket{f_{\bp, \bk, \lambda}} = \hat{d}^\dagger_{\bp - \bk\lambda} \hat{\beta}^\dagger_\bk \ket{\BEC}$. Here the dark state operator is defined as $\hat{d}^\dagger_{\bp\lambda} = \cos\theta_{\bp\lambda} \hat{\alpha}^\dagger_{\bp\lambda} - \sin\theta_{\bp\lambda} \hat{c}^\dagger_{\bp - \bk{\rm cl}}$, with $\tan \theta_{\bp\lambda}= \sqrt{n}g_{\bp\lambda} / \Omega$. Using Eq. \eqref{eq.H_eff} we then get
\begin{equation}
\Gamma_{\rm damp} = \pi\,\frac{n\T^2}{V}  \sum_{\bk,\lambda} (u_\bk - v_\bk)^2 \sin^2\theta_{\bp - \bk \lambda} \cdot \delta(\varepsilon_\bp - \varepsilon_{\bp - \bk} - \omega_{\bk}), \nn
\end{equation} 
using that $\sin\theta_{\bp\lambda_i} \simeq 1$. We let $\bq = \bp - \bk$ and let the dipole moment define the $z$-direction, $\hat{d}_{\rm be} = \bd_{\rm be} / |\bd_{\rm be}| = \hat{z}$. Then $g_{\bq\lambda} = g \, {\bm \epsilon}_{\bq \lambda}\cdot \hat{d}_{\rm be}$, with ${\bm \epsilon}_{\bq \lambda}$ the polarization vector. Since the polarizations have to be perpendicular to $\bq$ we can choose them as the spherical angle unit vectors: ${\bm \epsilon}_{\bq \lambda_1} = \hat{\theta}$, and ${\bm \epsilon}_{\bq \lambda_2} = \hat{\phi}$. Here $\theta, \phi$ are the polar and azimuthal angles respectively. Then $g_{\bq\lambda_1} = g \, \hat{\theta} \cdot \hat{z} = - g \sin\theta$, and $g_{\bq\lambda_2} = g \, \hat{\phi} \cdot \hat{z} = 0$. This in turn yields $\sin^2 \theta_{\bq \lambda} = ng_{\bq \lambda}^2 / (\Omega^2 + ng_{\bq \lambda}^2) = \delta_{\lambda, \lambda_1} ng^2 \sin^2\theta / (\Omega^2 + ng^2 \sin^2\theta)$. We then get for the damping rate
\begin{align}
\Gamma_{\rm damp}   &= \pi \, n\T^2   \int \frac{{\rm d}^3 q}{(2\pi)^3}  (u_{\bp - \bq} - v_{\bp - \bq})^2 \sin^2\theta_{\bq \lambda} \delta(\varepsilon_\bp - \varepsilon_{\bq} - \omega_{\bp - \bq}) \nn \\
                    &= \pi \, n\T^2   \int \frac{{\rm d}^3 q}{(2\pi)^3}  \frac{\xi_{\bp - \bq}}{\omega_{\bp - \bq}} \frac{ng^2 \sin^2\theta}{\Omega^2 + ng^2 \sin^2\theta} \delta(\varepsilon_\bp - \varepsilon_{\bq} - \omega_{\bp - \bq}), 
\label{eq.Gamma_damp_2}
\end{align}
using that $(u_{\bp - \bq} - v_{\bp - \bq})^2 = \xi_{\bp - \bq} / \omega_{\bp - \bq}$. In the energy difference of the $\delta$-function, we may approximate $\varepsilon_\bp - \varepsilon_{\bq} - \omega_{\bp - \bq} \simeq - v_{\rm g}(\theta) (q - p_0)$, evaluating the expression at the carrier momentum $\bp = \bp_0$. Inserting this in Eq. \eqref{eq.Gamma_damp_2} we then get
\begin{equation}
\Gamma_{\rm damp}   = \pi \, n\T^2   \int \frac{{\rm d}^3 q}{(2\pi)^3}  \frac{\xi_{\bp - \bq}}{\omega_{\bp - \bq}} \frac{ng^2 \sin^2\theta}{\Omega^2 c} \delta(q - p) \simeq \pi \, \frac{n\T^2}{(2\pi)^3 v_{\rm g}} p^2 \int_0^{2\pi} {\rm d}\phi \int_{0}^{\pi} {\rm d}\theta \frac{\xi_{\bp - \bq}}{\omega_{\bp - \bq}} \sin^3\theta,  
\label{eq.Gamma_damp_2}
\end{equation}
using that the group velocity for the dark state corresponding to the incoming photons is $v_{\rm g} = \Omega^2 / (\Omega^2 + ng^2) \cdot c \simeq \Omega^2 / ng^2 \cdot c$. Although this expression is rather difficult to evaluate analytically, we can come with a simple upper bound by approximating $\xi_{\bp - \bq} / \omega_{\bp - \bq} = 1$ -- it yields an upper bound since $\omega_{\bp - \bq} > \xi_{\bp - \bq}$. The integrals are then readily evaluated to $8\pi / 3$, and writing the damping rate in units of $\tB$ we finally get
\begin{equation}
\Gamma_{\rm damp} \simeq \frac{\sqrt{2}}{3} \frac{a^2}{\aB \xi}\frac{c_{\rm s}}{v_{\rm g}} \left(\frac{p}{p_{\rm c}}\right)^2,  
\label{eq.Gamma_damp_approx}
\end{equation}
with $p_{\rm c} = m c_{\rm s}$ the critical momentum of the BEC. This additional damping rate is thus significantly suppressed by $c_{\rm s} / v_{\rm g} \ll 1$. The factor of $\xi_{\bp - \bq} / \omega_{\bp - \bq}$ in the integrand can easily be included in a numerical calculation and leads to a further suppression of $\Gamma_{\rm damp}$ by about $10\%$ for the parameters used in Fig. 1(d), and by about $20\%$ for the parameters in Fig. 3. For completeness we show Fig. 1(d) of the main text corrected with this additional damping rate in Fig. \ref{fig.figS3}(a). Importantly, we see that $\Gamma_{\rm damp}$ is completely negligible for $\Omega > \gamma$ preserving the critical behaviour of the total scattering rate $\Gamma + \Gamma_{\rm damp}$. Further, in Fig. \ref{fig.figS3}(b) we plot the pulse propagation as in Fig. 3 of the main text. The dark state propagation including the additional damping rate calculated here is shown in green, and only shows a very small correction.  

\end{document}